\documentclass[useAMS,usenatbib,babel,times,onecolumn]{mn2e}
\usepackage{multirow,balance}
\usepackage[english,english]{babel}
\usepackage{amsmath}
\usepackage{amssymb,amsfonts,textcomp}
\usepackage{array}
\usepackage{supertabular}
\usepackage{hhline}
\usepackage{hyperref}
\usepackage[usenames]{color}
\hypersetup{dvips, colorlinks=true, linkcolor=black, citecolor=black, filecolor=black, urlcolor=black}
\usepackage[dvips]{graphicx}

\usepackage{graphicx}

\newcommand{\derd}{{\rm d}}
\newcommand{\be}{\begin{equation}}
\newcommand{\ee}{\end{equation}}
\newcommand{\ddir}{\delta_\text{D}}
\newcommand{\eh}[1]{\exp\left[#1\right]}
\newcommand{\ii}{\imath}

\def\ppk{P_{\rm pk}}

\def\xpk{\xi_{\rm pk}}
\def\gtrsim{\lower.5ex\hbox{$\; \buildrel > \over \sim \;$}}
\def\bnpk{\bar{n}_{\rm pk}}
\def\npk{n_{\rm pk}}
\def\dpk{\delta_{\rm pk}}
\def\la{\langle}
\def\ra{\rangle}

\definecolor{grey}{rgb}{0.75,0.75,0.75}
\definecolor{Orange}{rgb}{1.0,0.5,0.15}
\definecolor{brown}{rgb}{0.7,0.25,0.0}
\definecolor{Pink}{rgb}{1.0,0.5,0.5}
\definecolor{darkerred}{rgb}{0.8,0,0}
\definecolor{darkerblue}{rgb}{0,0,0.8}
\definecolor{Blue}{rgb}{0,0.08,0.65}
\definecolor{Red}{rgb}{0.65,0.08,0.05}
\definecolor{Green}{rgb}{0.15,0.45,0.25}

\begin{document}
\author[Baldauf et al.]{
Tobias Baldauf$^{1}$, Sandrine Codis$^{2,3}$, Vincent Desjacques$^{4}$ and Christophe Pichon$^{2,5}$
\vspace*{6pt}\\
\noindent $^{1}$ Institute for Advanced Study, School of Natural Sciences, Einstein Drive, 
Princeton, NJ 08540, USA\\
$^{2}$ CNRS, UMR7095 \& UPMC  Institut d'Astrophysique de Paris, 
98 bis Boulevard Arago, 75014, Paris, France\\
$^{3}$ Canadian Institute for Theoretical Astrophysics, University of Toronto, 60 St. George Street, Toronto, ON M5S 3H8, Canada \\
$^{4}$ D\'epartement de Physique Th\'eorique, Universit\'e de Gen\`eve, 24, quai Ernest Ansermet. 
1211, Gen{\`e}ve, Switzerland\\
$^{5}$ Institute of Astronomy, University of Cambridge, Madingley Road, Cambridge, CB3 0HA, United Kingdom\\
}

\title[Peak exclusion in 1+1 gravity]
{Peak exclusion, stochasticity and 
convergence of perturbative bias expansions in 1+1 gravity}

\maketitle

\begin{abstract}
The Lagrangian peaks of a 1D cosmological random field representing dark matter  are used as a 
proxy for a catalogue of biased tracers in order to investigate 
the small-scale exclusion in the two-halo term.  
The two-point correlation function of peaks of a given height is numerically estimated and analytical approximations that are valid inside the exclusion zone are derived. The resulting power spectrum of these tracers is investigated and shows clear deviations from Poisson noise at low frequencies.
On large scales, the convergence of a perturbative bias expansion is discussed. 
Finally, we  go beyond  Gaussian statistics for the initial conditions and investigate the 
subsequent evolution of the two-point clustering of peaks through their Zel'dovich ballistic displacement,
  to clarify how exclusion effects mix up with scale-dependencies
induced by nonlinear gravitational evolution. While the expected large-scale separation limit is recovered, significant deviations are found in the exclusion zone that tends in particular to be reduced at later times.
Even though these findings apply to the clustering of one-dimensional tracers, 
they provide useful insights into halo exclusion and its impact on the two-halo term.

\end{abstract}


%%%%%%%%%%%%%%%%%%%%%%%%%%%%%%%%%%%%%%%%%%%%%%%%%%%%%%%%%%%%%%%%%%%%%%%%%%
\section{Introduction}
%%%%%%%%%%%%%%%%%%%%%%%%%%%%%%%%%%%%%%%%%%%%%%%%%%%%%%%%%%%%%%%%%%%%%%%%%%

Dark matter haloes and the galaxies within them are distinct and extended objects. By definition, 
they cannot overlap since their centres have to be separated by at least the sum of their virial 
radii. This exclusion effect is even more important in the initial conditions or Lagrangian space, 
before the objects collapsed and fell towards each other. 
As noted in \cite{mo/white:1996,sheth/lemson:1999}, the vanishing probability to find two centres 
closer than the exclusion radius corresponds to the correlation function being $-1$ for small 
separations. While this effect is localized at small separations in the correlation function, 
it can alter the power spectrum at small wavenumbers (large scales) and results in a modification of Poisson 
stochasticity \citep{mo/white:1996,sheth/lemson:1999,smith/scoccimarro/sheth:2007,baldauf:2013excl} 
consistent with the sub-Poissonian noise measured in the clustering of simulated dark matter haloes
\citep{casasmiranda/mo/etal:2002,seljak/hamaus/desjacques:2009,hamaus/seljak/etal:2010,manera/gaztanaga:2011}.
In addition, exclusion effects strongly suppress the non-physical $k^0$ tail of the one-halo term
\citep{smith/desjacques/marian:2011}. 

Besides the exclusion effects, there are distinct, non-linear bias effects just outside the exclusion 
region which, due to their localization, also contribute to the power spectrum on large scales and 
for which the bias expansion converges very slowly. 
The exclusion region and the non-linear bias bump beyond are important for precision models of the 
halo-halo correlation function in the transition region between the one- and two-halo terms in the 
halo model. Thus, a better understanding of these regions will likely improve the modelling of the 
matter power spectrum or correlation function in this regime, which is very important for weak 
lensing or galaxy-galaxy lensing studies, since this signal is large and not yet dominated by the 
fully non-perturbative one-halo term.

Dark matter haloes are seeded by over-dense regions in Lagrangian space (so called proto-haloes) that 
subsequently collapse to form the virialized late-time Eulerian haloes. Various assumptions can be
 made to describe the relation between the proto-haloes and the underlying Gaussian density field. 
Our perfect knowledge of the $N$-point statistics of the Gaussian field allows us to calculate all 
possible statistics of transformations of the Gaussian field. In this paper we will consider the 
peak model \citep{bardeen/bond/etal:1986,Regos95}, in which proto-haloes are associated with the maxima of the 
smoothed underlying field.
To simplify the calculations and understanding, but without loosing much of the phenomenology, we 
will consider volume exclusion effects associated with peaks in one spatial dimension, following 
\cite{lumsden/heavens/peacock:1989,coles:1989}. The 1D approach keeps the calculation simple, reducing the
number of field variables to be considered at each point from 10 in 3D to 3 in 1D.

The paper is organized as follows. In Section~\ref{sec:montecarlo}, the general mathematical formalism that 
defines peak-peak correlations functions in dimension $d$, together with their numerical implementation are described. 
Section~\ref{sec:exclusion} presents the result on the small-scale exclusion zone of peaks obtained by 
numerical integration. 
The analytical large-scale bias expansion of the two-point correlation function of peaks is then discussed 
in Section~\ref{sec:bias}. 
Section~\ref{sec:zeldo} incorporates the effect of the Zel'dovich displacement of the peaks and their velocity statistics.
Finally, Section~\ref{sec:conclusion} wraps up.

%%%%%%%%%%%%%%%%%%%%%%%%%%%%%%%%%%%%%%%%%%%%%%%%%%%%%%%%%%%%%%%%%%%%%%%%%%
\section{Formalism and numerical implementation}
%%%%%%%%%%%%%%%%%%%%%%%%%%%%%%%%%%%%%%%%%%%%%%%%%%%%%%%%%%%%%%%%%%%%%%%%%%
\label{sec:montecarlo}

The formalism of cosmological density peaks, which builds on the Kac-Rice formula 
\cite{Kac1943,Rice1945} was laid down in \cite{bardeen/bond/etal:1986}. Following 
\cite{pogosyan/pichon/etal:2009}, in $d$ dimensions,  for a given (overdensity) field $\rho$, we define 
the moments
\begin{align}
{\sigma_0}^2 &= \langle \rho^2 \rangle, 
& {\sigma_1}^2 &= \langle \left( \nabla \rho \right)^2 \rangle, 
& {\sigma_2}^2 &= \langle (\Delta \rho)^2 \rangle.
\end{align}
Combining these moments, we can build two characteristic lengths
$R_0 ={\sigma_0}/{\sigma_1}$ and $R_\star = {\sigma_1}/{\sigma_2}$, as well as the 
spectral parameter
\begin{equation} 
\gamma=\frac{{\sigma_1}^2}{\sigma_0 \sigma_2}.
\end{equation}
We choose to normalise the field and its derivatives to have unit variances:
\begin{align} 
x&=\frac{1}{\sigma_0} \rho, 
& x_i&=\frac{1}{\sigma_1} \nabla_i \rho, 
& x_{ij} &=\frac{1}{\sigma_2} \nabla_i \nabla_j \rho.
\end{align}
In general, while ${\cal P}(\boldsymbol{X})$ designates the one-point probability density (PDF), 
${\cal P}(\boldsymbol{X},\boldsymbol{Y})$ will denote the joint PDF for the normalized 
field and its derivatives, $\boldsymbol{X}=\{x,x_{ij},x_i\}$ and $\boldsymbol{Y}=\{y,y_{ij},y_i\}$, 
at two prescribed comoving locations (${\boldsymbol r}_{x}$ and ${\boldsymbol r}_{y}$ separated by a distance 
$r=|{\boldsymbol r}_{x}-{\boldsymbol r}_{y}|$). In the particular case of Gaussian initial conditions, this 
joint PDF is the multivariate Normal
\begin{equation}
{\cal N}(\boldsymbol{X},\boldsymbol{Y})= \frac{\exp\left[-\frac{1}{2}
\left(\begin{array}{c}
 \boldsymbol{X} 
\\
\boldsymbol{Y} 
 \\
\end{array} \right)^{\rm T}
 \cdot
  \mathbf{C}
 ^{-1}\cdot \left(\begin{array}{c}
 \boldsymbol{X} 
\\
\boldsymbol{Y} 
 \\
\end{array} \right) \right]}{{\rm det}|\mathbf{C}|^{1/2} \left(2\pi\right)^{\rm (d+1)(d+2)/2 }} \,, 
\label{eq:defPDF}
\end{equation} 
where $\mathbf{C}_{0}\equiv \langle  \boldsymbol{X}\cdot \boldsymbol{X}^{\rm T} \rangle$ and  
$\mathbf{C}_{\gamma}\equiv \langle  \boldsymbol{X}\cdot \boldsymbol{Y}^{\rm T} \rangle$ are the diagonal
and off-diagonal components of the covariance matrix
\begin{equation}
\quad \mathbf{C}=\left(
\begin{array}{cc}
\mathbf{C}_{0} &\mathbf{C}_\gamma
\\
\mathbf{C}_\gamma^{\rm T}  &\mathbf{C}_{0} 
 \\
\end{array}
\right)\,.
\end{equation}
All these quantities depend on the separation vector $\boldsymbol{r}$ only because of homogeneity. 
Isotropy further implies that they depend on the modulus $r=|\boldsymbol{r}|$ solely.
Equation~(\ref{eq:defPDF}) is sufficient to compute the expectation of any quantity involving the 
fields and its derivatives up to second order.
In particular, the two-point correlation $\xi_\text{crit}(r,\nu)$ of (signed) critical points 
at threshold $\nu$ separated by $r$ is given by
\begin{equation}
\label{eq:xicrit}
1+\xi_{\rm crit}(r,\nu)=
 \frac{
\big\langle n_{\rm crit}(\boldsymbol{X})  n_{\rm crit}(\boldsymbol{Y}) \big\rangle}
{\big\langle n_{\rm crit}(\boldsymbol{X}) \big\rangle^2}   \,,
\end{equation}
where the Klimontovich or ``localized'' density for a signed critical point reads 
\begin{equation}
\label{eq:ncrit}
n_{\rm crit}(\boldsymbol{X})=\left(\frac{\sigma_2}{\sigma_1}\right)^{d}
{\rm det}(x_{ij})\delta_{\rm D}(x_i)\delta_{\rm D}(x-\nu) \,.
\end{equation}
This density is formally zero unless the condition for a critical point is satisfied. The 
multiplicative factor of $(\sigma_1/\sigma_2)^{d}$, which has dimension of 
length$^{-{d}}$, ensures that the ensemble average 
\begin{equation}
\big\langle n_{\rm crit}(\boldsymbol{X})  \big\rangle
=  \int\!{\rm d} \boldsymbol{X} \,
  {\rm det}(x_{ij})  \delta_{\rm D}(x_i)  \delta_{\rm D}(x-\nu)     {\cal P}(\boldsymbol{X})
\equiv \bar{n}_\text{crit}(\nu)\,,
\end{equation}
which appears in the denominator of equation~(\ref{eq:xicrit}), equals the average number density 
of critical points at threshold $\nu$. The ensemble average
\begin{equation}
\big\langle n_{\rm crit}(\boldsymbol{X}) n_{\rm crit}(\boldsymbol{Y})\big\rangle 
=  \int\!{\rm d}\boldsymbol{X}\!\int\!{\rm d}\boldsymbol{Y} \, {\cal P}(\boldsymbol{X},\boldsymbol{Y}) \,
{\rm det}(x_{ij})  \delta_{\rm D}(x_i)  \delta_{\rm D}(x-\nu)
{\rm det}(y_{ij})  \delta_{\rm D}(y_i)  \delta_{\rm D}(y-\nu) 
\end{equation}
is the cross-correlation.
Since the integrand  is simply a polynomial function of the variables, this integral can be 
fully carried out analytically.
For peaks, an additional constraint on the sign of the second derivatives is required. As a 
consequence, the peak two-point correlation becomes
\begin{equation}
\label{eq:xipeak}
1+\xpk(r,\nu)= 
 \frac{\big\langle n_{\rm pk}(\boldsymbol{X}) n_{\rm pk}(\boldsymbol{Y})\big\rangle}
{\big\langle n_{\rm pk}(\boldsymbol{X})\big\rangle^2} \,.
\end{equation}
where the localized peak number density $n_{\rm pk}(\boldsymbol{X})$, 
\begin{equation}
\label{eq:npk}
n_{\rm pk}(\boldsymbol{X}) =
\left(\frac{\sigma_2}{\sigma_1}\right)^{d}
|{\rm det}(x_{ij})|  \delta_{\rm D}(x_i)  \Theta_{\rm H}(-\lambda_{i})  \delta_{\rm D}(x-\nu) \,,
\end{equation}
implements the peak condition.
In odd dimensions (e.g $d=1$), $ |{\rm det}(x_{ij})|=- {\rm det}(x_{ij})$ because the determinant is negative at the 
peaks and it is understood that, for $d>1$,  $\delta_{\rm d}(x_i)$ stands for  
$\prod_{i\le d} \delta_{\rm D}(x_i)$,
while  $ \Theta_{\rm H}(-y_{ii}) $  means $\prod_{l\le d} \Theta_{\rm H}(-\lambda_l)$, with 
$\{\lambda_l\}_l$ being the eigenvalues of the Hessian.
Because of these inequalities, the integral typically is not analytical anymore.

{  It has to be noted that contrary to peaks, the number density of signed critical points is not restricted to be positive. In particular, for Gaussian statistics, the number density of peaks and minima are related via $n_{\rm pk}(\nu)=n_{\rm min}(-\nu)$ so that $n_{\rm crit}(\nu)$ { (which is nothing but the alternating sum of minima and peaks $n_{\rm min}(\nu)-n_{\rm pk}(\nu)$)} is positive for $\nu<0$ and negative for $\nu>0$. In particular, $\xi_{\rm crit}$ is then expected to diverge for $\nu=0$ and unlike peaks, it is not restricted to be above $-1$.}

In $d$ dimension, we define the conditional probability that $x_{ij}$ and $y_{ij}$ 
satisfy the PDF, subject to the condition that $x_i=y_i=0$ and $x=y=\nu$ and 
resort to Monte-Carlo 
methods in {\small MATHEMATICA} in order to evaluate numerically equation~(\ref{eq:xipeak}). Namely, we draw random 
numbers of dimension 
$d(d+1)$ from the conditional probability that $x_{ij}$ and $y_{ij}$ satisfy 
the PDF, subject to the condition that $x_i=0$ and $x=y=\nu$ (using {\tt RandomVariate}). 
For each draw $^{(k)}$ if $\lambda_l(x^{(k)}_{ij})<0 $ and  $\lambda_l(y^{(k)}_{ij})<0 $ 
($l\le d$) we keep the sample and evaluate $ {\rm det}(x^{(k)}_{ij})   {\rm det}(y^{(k)}_{ij}) $ 
and otherwise we drop it; eventually, 
\begin{equation}
\big\langle n_{\rm pk}(\boldsymbol{X})n_{\rm pk}(\boldsymbol{Y})\big\rangle
 \approx \frac{1}{N}
\sum_{k\in {\cal S}} \left[    {\rm det}(x^{(k)}_{ij})  {\rm det}(y^{(k)}_{ij})\right]\times {\cal P}(x=y=\nu,x_{i}=y_{i}=0) \,,
\end{equation}
where $N$ is the total number of draws, and $\cal S$ is the subset of the indexes of draws 
satisfying the constraints on the eigenvalues.
The same procedure can be applied to evaluate the denominator
$\left\langle n_\text{pk}(\boldsymbol{X}) \right\rangle\equiv \bnpk(\nu)$. 
Equation~(\ref{eq:xipeak}) then yields  $\xpk(r,\nu)$. This algorithm is embarrassingly 
parallel and can be easily generalized, for instance, to the computation of the correlation 
function $\xpk(r,>\nu)$ of peaks {\sl above} a given threshold in density and to arbitrary dimension $d$. 
In practice it is fairly efficient as the draw is customized to the shape of the underlying 
Gaussian PDF. For $d=1$ considered in this work, this brute force Monte-Carlo method converges 
relatively quickly. Namely, we use one million draws for each evaluation of the correlation function 
in this work. This number is sufficient to reach percent precision accuracy. 
Obviously, if correlation function above a given threshold are considered, the required number of draws 
is larger and increases with the value of the threshold (as the event $x>\nu$ becomes rarer).

%%%%%%%%%%%%%%%%%%%%%%%%%%%%%%%%%%%%%%%%%%%%%%%%%%%%%%%%%%%%%%%%%%%%%%%%%%
\section{Small scales: peak-peak exclusion}
%%%%%%%%%%%%%%%%%%%%%%%%%%%%%%%%%%%%%%%%%%%%%%%%%%%%%%%%%%%%%%%%%%%%%%%%%%
\label{sec:exclusion}
Here and henceforth, we will assume Gaussian initial conditions such that 
${\cal P}(\boldsymbol{X},\boldsymbol{Y})$=${\cal N}(\boldsymbol{X},\boldsymbol{Y})$. 
In 1D, the block matrices that make up the covariance matrix are
\begin{equation}
\mathbf{C}_0=\left(
\begin{array}{ccc}
 1 & 0&-\gamma   \\
0 & 1 & 0 \\
 -\gamma  & 0 & 1 \\
\end{array}
\right) \,,\quad \mathbf{C}_\gamma=\left(
\begin{array}{ccc}
 \gamma _{00} & \gamma _{01} &
   \gamma _{02} \\
 \gamma _{01} & \gamma _{11} &
   \gamma _{12} \\
 \gamma _{02} & \gamma _{12} &
   \gamma _{22} \\
\end{array}
\right) \,,
\end{equation}
where the $\gamma_{ ij}$'s represent the correlations between the field and its derivatives 
at two points separated by a comoving distance $r$, 
e.g. $\gamma_{22}=\langle  x_{11} y_{11} \rangle$. 
These $\gamma_{ ij}$'s are not independent. The following relations are established via 
integrations by part: 
$\gamma_{10}=-\gamma_{01},\quad\gamma_{21}=-\gamma_{12},\quad \gamma_{20}=-\gamma \gamma_{11}$.
%>>>>>>>>>>>>>>>>>>>>>>>>>>>>>>>>>>>>>>>>>>>>>>>>>>>>>>>>>>>>>>>>>>>>>>>>>>>>>>>>>>>>
\begin{figure*}
\centering
\includegraphics[width=0.45\columnwidth]{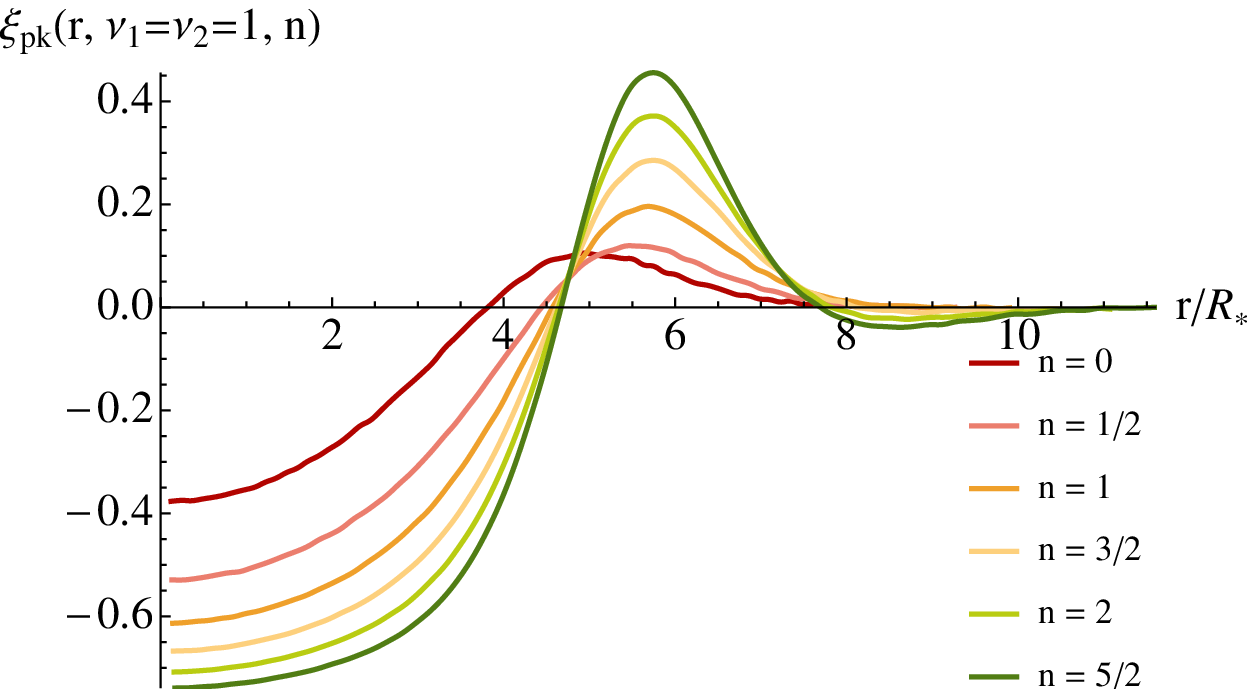} 
\includegraphics[width=0.45\columnwidth]{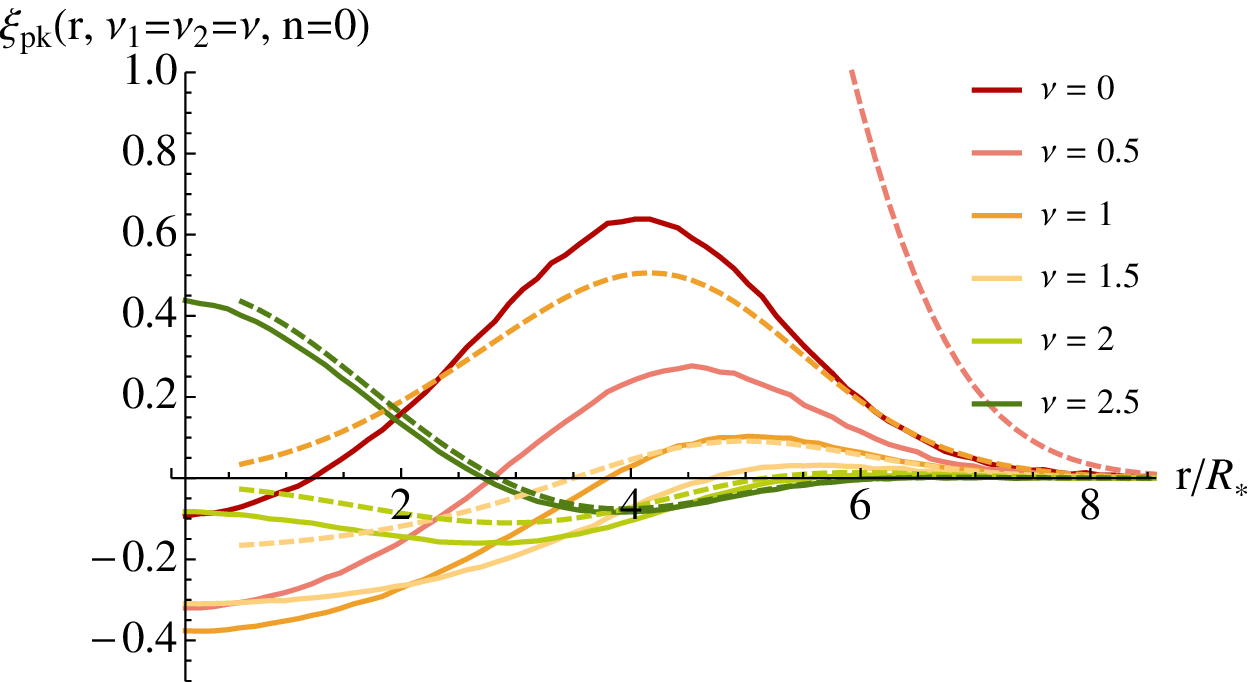}
\caption{{\it Left panel}: 1D correlation function of peaks of height $\nu=1$ as a function of 
$r/R_\star$ for different spectral index $n$ evaluated by Monte-Carlo realisations of the peak 
constraint. {\it Right panel}: Same as left panel, but the
spectral index is held fixed at $n=0$ while the peak height $\nu$ is varied between 0 and 2.5
as labelled. The dashed lines indicate the analytical 1D signed critical point correlation  
function. For $\nu> 2$, both the peak and signed critical points correlations are in very good 
agreement. Note that, for $\nu=0$, the signed critical point correlation functions diverges {  (because the alternating sum of $\nu=0$ critical points goes to zero)} and 
is thus not shown on this plot. 
\label{fig:1Dpeakcrit}}
\end{figure*}
%>>>>>>>>>>>>>>>>>>>>>>>>>>>>>>>>>>>>>>>>>>>>>>>>>>>>>>>>>>>>>>>>>>>>>>>>>>>>>>>>>>>>
The  $\gamma_{ij}( r)$ are known function of $r$ given by the moments of the two fields and 
their derivatives:
\begin{equation}
\label{eq:gammaij}
\gamma_{ij}(r) =
\frac{1}{ \sigma_i\sigma_j }\int {\rm d}k\, \exp(\imath k r) (\imath k)^{i}(-\imath k)^{j} P_s(k)
\quad i\leq j\,,
\end{equation}
with $P_s(k)$ the power spectrum of $\delta$ smoothed with a filter function (assumed Gaussian 
throughout this work). On expanding $\gamma_{ij}( r)$ at small separations $r\ll R$ and 
substituting the spectral moments 
\begin{equation}
\label{eq:sigma}
\sigma_{l}^{2}=\frac{1}{\pi}\int_{0}^{\infty}{\rm d}k\, k^{2l} P_s(k)\,,
\end{equation}
it follows that
\begin{align}
\gamma_{ij}(r)& \!=\!
\frac{(-1)^{l-i}}{ \sigma_i\sigma_j }\!\sum_{k=0}^{\infty}\frac{(-1)^{k}}{(2k)!}\!\left(\frac{r}{R}\right)^{2k}
\sigma_{l+k}^{2} \qquad\qquad\quad\;\; (i+j=2l)\;, \\
\gamma_{ij}(r)& \!=\!
\frac{(-1)^{1+l-i}}{ \sigma_i\sigma_j }\!\sum_{k=0}^{\infty}\frac{(-1)^{k}}{(2k+1)!}\!\left(\frac{r}{R}\right)^{2k+1}
\sigma_{l+k+1}^{2} \quad\;(i+j=2l+1)\;,
\end{align}
where $\tilde r=r/R$ is the separation in units of the smoothing length $R$. 
The determinant of the covariance matrix $\mathbf{C}$ is given 
at {  leading} order in the separation $r$  by $(r/R)^{18}\times g(\{\sigma_{i}\}_{0\geq i\geq 5})$ where 
$g= \left(\sigma _2^6-\left(2 \sigma _1^2 \sigma _3^2+\sigma _0^2 \sigma _4^2\right) \sigma _2^2
+\sigma _0^2 \sigma_3^4+\sigma _1^4 \sigma _4^2\right)\!\times $ \\
$\left(\sigma _3^6-\left(2 \sigma _2^2 \sigma _4^2+\sigma _1^2 \sigma _5^2\right)
   \sigma _3^2+\sigma _1^2 \sigma _4^4+\sigma _2^4 \sigma _5^2\right)/$$74649600 \sigma _0^4 \sigma _1^4 \sigma _2^4$
does not depend on the separation.
Indeed, three eigenvalues of $\mathbf{C}$ are singular, respectively scaling like $ r^{10}$, $ r^{6}$ 
and $ r^{2}$ and corresponding to the eigen-directions given by $(x-y)$, $(x_{1}-y_{1})$ and 
$(x_{11}-y_{11})$. This singularity proportional to $ r^{-18}$ is the reason why the limit 
$ r\rightarrow 0$ is difficult to handle numerically. Analytically, this means that a series expansion 
to eighteenth order is needed for all terms. {  Note that \cite{lumsden/heavens/peacock:1989} only expand the covariance matrix to second order. Truncating the expansion at this order we find that the determinant scales as $r^{10}$ and the 11 element of the inverse as $\mathbf{C}_{11}^{-1}\propto r^{-4}$. Their expansion was thus not sufficient to account for the cancellations between terms at small separations. \footnote{{ Note that this difficulty arises only in the low-$r$ behaviour and therefore does not  affect the full computations done in that paper.}}}

%-------------------------------------------------------------------------
\subsection{Correlation of 1D peaks of same height}
%-------------------------------------------------------------------------

We evaluate the two-point correlation function of $d=1$ peaks upon applying the Monte-Carlo method 
described above to equation~(\ref{eq:xipeak}). Results {  for a power-law power spectrum $P_s(k)= A k^{n}$} are shown in the left panel of 
Fig.~\ref{fig:1Dpeakcrit} as a function of the spectral index $n$ for a fixed peak height $\nu=1$. 
The exclusion zone shrinks to smaller separations and becomes more pronounced as $n$ is increased 
because the addition of small-scale power tends to sharpen the profile around local density maxima. 
In the right panel of Fig.~\ref{fig:1Dpeakcrit}, we display $\xpk(r,\nu)$ as a function of peak 
height for a white noise power spectrum $n=0$. 
For comparison, the dashed curves represent the two-point
correlation of 1D signed critical points, which is obtained upon integrating equation~(\ref{eq:xicrit}) 
over the six field variables. 
Unsurprisingly, $\xi_\text{crit}(r,\nu)$ matches $\xpk(r,\nu)$ 
almost perfectly for prominent peaks ($\nu\gtrsim 2$) since, in this regime, a critical point is 
nearly always a local maximum. 

Interestingly, the two-point correlation $\xi_\text{crit}(r,\nu)$ is fully analytical regardless of 
the underlying density power spectrum. For the sake of readability however, we will not display 
its full expression here. Nevertheless, we can take advantage of this analytic result, together with 
the fact that $\xpk(r,\nu)$ agrees very well with its genus-like counterpart $\xi_\text{crit}(r,\nu)$
at high threshold, to get insights into the short distance behaviour of the peak correlation function. 
The low-$r$ limit of the two-point correlation function of signed critical points separated by $r$ 
is given by
\begin{equation}
1+\xi_\text{crit}( r,\nu)
=\frac{e^{\frac{\left(2 \gamma ^2-1\right) \nu ^2}{2 \left(\gamma ^2-1\right)}} 
\left(\gamma _{\#}^2 \left(\gamma ^2 \gamma _\star ^4
\left(\gamma ^2+\nu ^2-1\right)-2 \gamma ^2 \gamma _\star ^2 \left(\gamma ^2+\nu ^2-1\right)+\gamma ^2 
\left(\nu^2+1\right)-1\right)+\left(\gamma ^2-1\right)^2\right)}{12 \gamma ^3 \left(1-\gamma ^2\right)^{5/2} 
\nu ^2 \gamma _{\#}^2   \sqrt{1-{\gamma _\star ^2}} \gamma _\star ^3}+{\cal O}( r/R)
\end{equation}
where we define the following  extra shape parameters 
$\gamma_\star=\sigma_{2}^{2}/\sigma_{1}/\sigma_{3}$ and 
$\gamma_{\#}=\sigma_{3}^{2}/\sigma_{2}/\sigma_{4}$.

%>>>>>>>>>>>>>>>>>>>>>>>>>>>>>>>>>>>>>>>>>>>>>>>>>>>>>>>>>>>>>>>>>>>>>>>>>>>>>>>>>>>>
\begin{figure*}
\centering
\includegraphics[width=0.45\columnwidth]{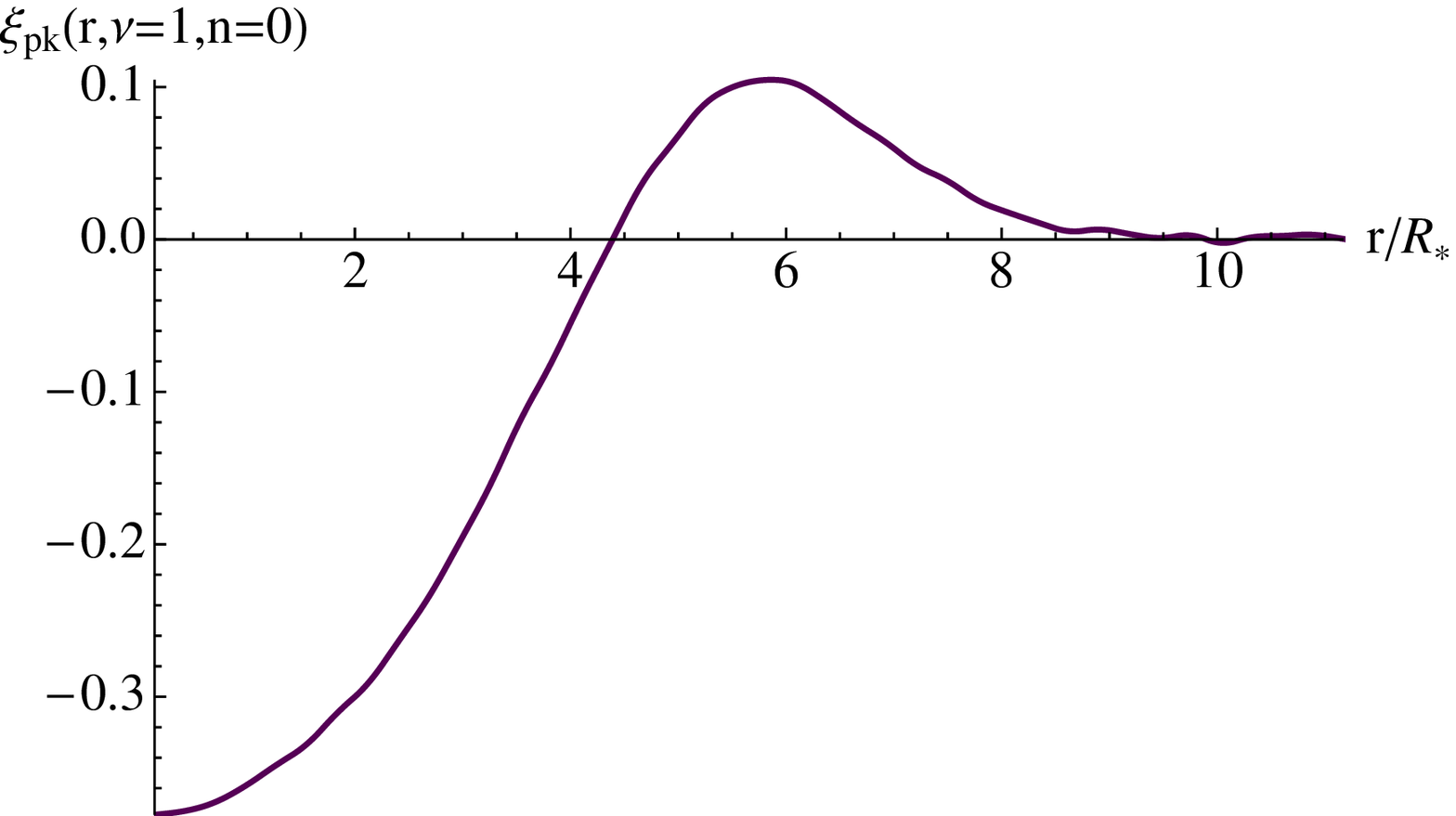} 
\includegraphics[width=0.45\columnwidth]{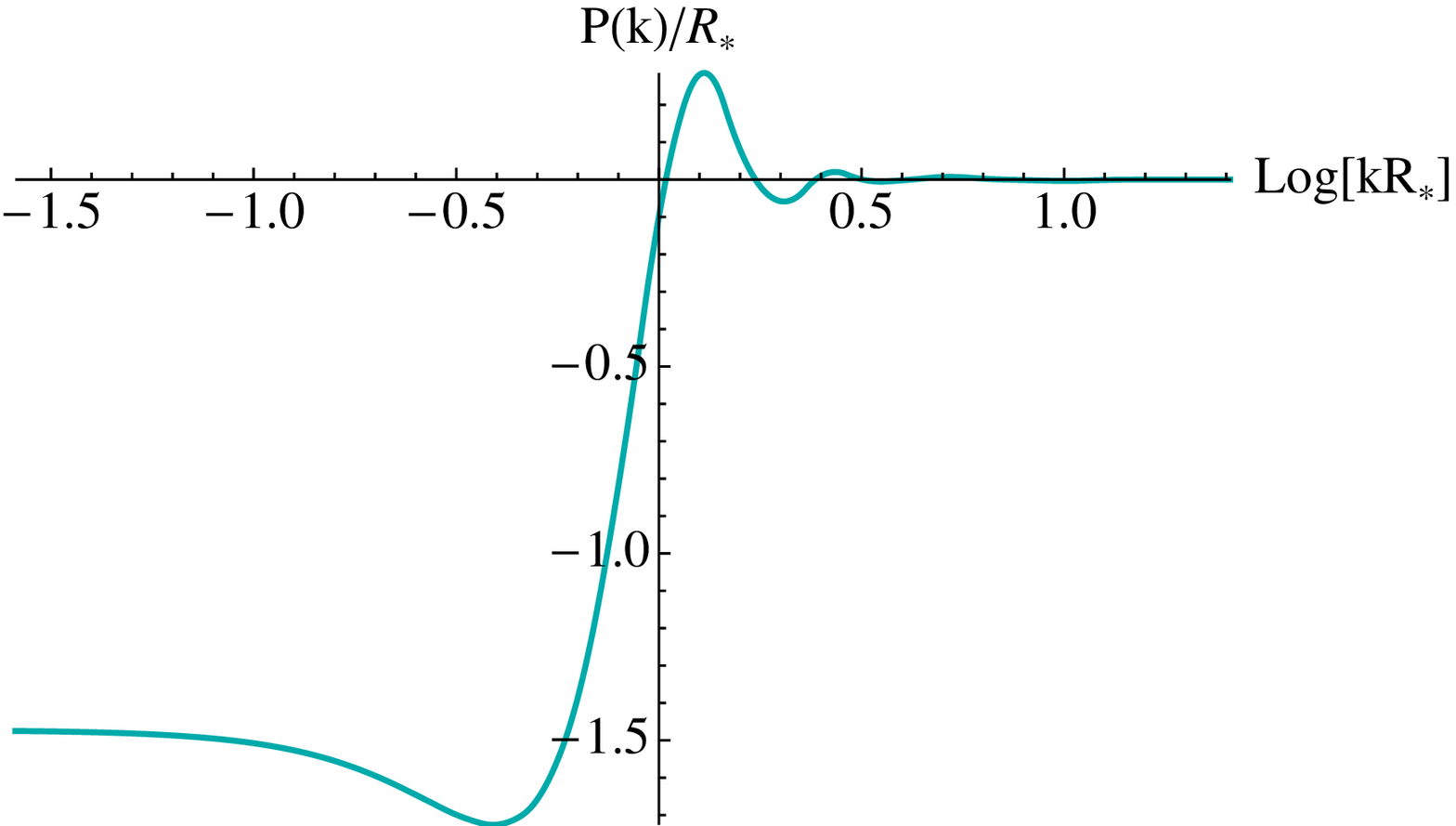}
\caption{{\it Left panel}: Monte-Carlo estimation of the two-point correlation function of 1D peaks 
$\xpk(r,\nu)$. The results is shown as a function of $r/R_\star$ assuming $n=0$ and a threshold 
$\nu=1$. {\it Right panel}: Corresponding power spectrum $\ppk(k,\nu)$. The exclusion zone seen at
short distance in the left panel makes the 1D peak power spectrum sub-Poissonian at small wavenumber
(large separation).\label{fig:zoomexclusion}}
\end{figure*}
%>>>>>>>>>>>>>>>>>>>>>>>>>>>>>>>>>>>>>>>>>>>>>>>>>>>>>>>>>>>>>>>>>>>>>>>>>>>>>>>>>>>>

For a power-law power spectrum $P_s(k)= A k^{n}$, with spectral index $n>-1$, and a density field 
filtered with a Gaussian kernel of radius $R$, the $\gamma_{ij}(r)$ are given by 
\begin{align}
\gamma _{00}( r)&= \, _1F_1\left(\frac{n+1}{2};\frac{1}{2};-\frac{ r^{2}}{4 R^{2}} \right)\,,\\
 \gamma _{11}( r)&= \, _1F_1\left(\frac{n+3}{2};\frac{1}{2}; -\frac{ r^{2}}{4 R^{2}} \right)\,, \nonumber \\
  \gamma_{22}( r)&=\, _1F_1\left(\frac{n+5}{2};\frac{1}{2}; -\frac{ r^{2}}{4 R^{2}} \right)\ \nonumber ,
\end{align}
and $\gamma _{02}( r)= -\gamma \gamma_{11}(r)$. Here, $_1 F_1$ is the Kummer confluent 
hypergeometric function and
\begin{equation}
\sigma_{l}^2\propto R^{-1-n-2l}\,\Gamma\!\left(\frac{1}{2}+\frac{n}{2}+l\right)\;.
\end{equation}
Therefore, the shape parameters are $\gamma=(1+n)/(3+n)$, $\gamma_\star=(3+n)/(5+n)$ and 
$\gamma_{\#}=(5+n)/(7+n)$, and the determinant of $\mathbf{C}$ thus scales like 
${\gamma ^2 \left(2-\gamma ^2\right)\left(1-\gamma ^2\right)^{-3}( r/R)^{18}}/{18662400 }$, as 
advertised at the beginning of this section. 
Note that, in this work, all the figures display the separation in units of $\tilde r =r/R_{\star}$ instead of the smoothing length $R$ as $R_{\star}$ represents the typical distance between extrema \citep{bardeen/bond/etal:1986} and is therefore more meaningful. 
In the case where $n=0$, the low-$r$ behaviour of the two-point correlation function of signed critical points can be written as follows
\begin{equation}
   1+\xi_\text{crit}(\tilde r,\nu,n=0)=
\frac{e^{\frac{\nu ^2}{4}} \left(3 \nu ^2+8\right)}{8 \sqrt{3} \nu ^2}+\frac{e^{\frac{\nu ^2}{4}} 
\left(128-15 \nu ^4\right) \tilde r^2}{1920 \sqrt{3} \nu   ^2}
+\frac{e^{\frac{\nu ^2}{4}} \left(15 \nu ^4-64\right)\tilde  r^4}{184320 \sqrt{3}}+{\cal O}\left(\tilde r^5\right)\;,
\label{eq:lowrequal}
\end{equation}
which makes clear that $\xi_\text{crit}$ diverges for $\nu=0$. 
The dependence of the low-$r$ expansion with the spectral index $n$ for peaks of height $\nu=1$ reads
\begin{equation}
   1+\xi_\text{crit}(\tilde r,\nu=1,n)
=\!\frac{e^{\frac{1}{4}-\frac{n}{4}} (n+3) \left(n^2+4 n+11\right)}{24 (n+1) \sqrt{n^2+4 n+3}}
+\frac{e^{\frac{1}{4}-\frac{n}{4}} (n+3) \left(-25 n^4+40 n^3+410   n^2+1464 n
+1695\right) \tilde r^2}{86400 (n+1) \sqrt{n^2+4 n+3}}+{\cal O}\left(\tilde r^3\right),
\end{equation}
which shows that the exclusion zone is more pronounced for high values of the spectral index. 
The same trend was also seen for peaks in the left panel of Fig.~\ref{fig:1Dpeakcrit}.
As shown in \cite{baldauf:2013excl}, the small-scale peak repulsion in the configuration space 
correlation function can have a significant impact on the power spectrum at small wavenumbers. 
To emphasize this point, Fig.~\ref{fig:zoomexclusion} displays, in the left panel, the 1D peak 
correlation function as a function of $\tilde r =r/R_{ \star}$ assuming $n=0$ and a threshold $\nu=1$ 
and, in the right panel, the corresponding power spectrum obtained by a simple Fourier transform
of the real space Monte-Carlo result
\begin{equation}
P_{\text{pk}}(k) = 2\int_0^\infty\!\!{\rm d}r\, \xi_{\text{pk}}(r)\cos (k r)\;.
\end{equation}
 In this particular case, the power spectrum is approximately
white for all wavenumbers {  $k\lesssim 1/R_\star$}, with $\ppk(k\lesssim R_\star,\nu=1)\approx -1.5 R_{\star}$.

%-------------------------------------------------------------------------
\subsection{How strong is the small-scale exclusion?}
\begin{figure}
\centering
\includegraphics[width=0.45\columnwidth]{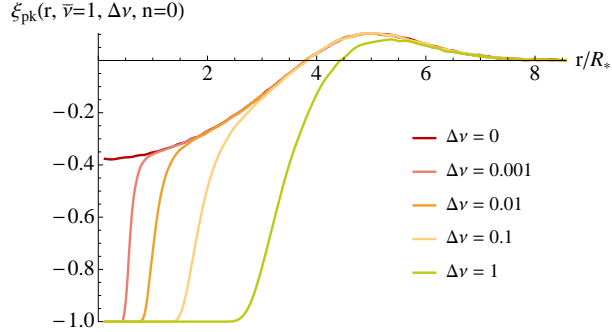}
\caption{The two-point correlation function of 1D peaks with height $\nu_{1}=1-\Delta \nu/2$ and 
$\nu_{2}=1+\Delta \nu/2$ is shown as a function of the height difference $\Delta\nu$ as labelled 
in the Figure. A fixed value of $\bar{\nu}\equiv (\nu_1+\nu_2)/2=1$ was assumed. All the correlation 
functions were estimated using the Monte-Carlo method described in Section~\ref{sec:montecarlo}.
\label{fig:varyingnu}}
\end{figure}

\begin{figure}
\centering
\includegraphics[width=0.49\columnwidth]{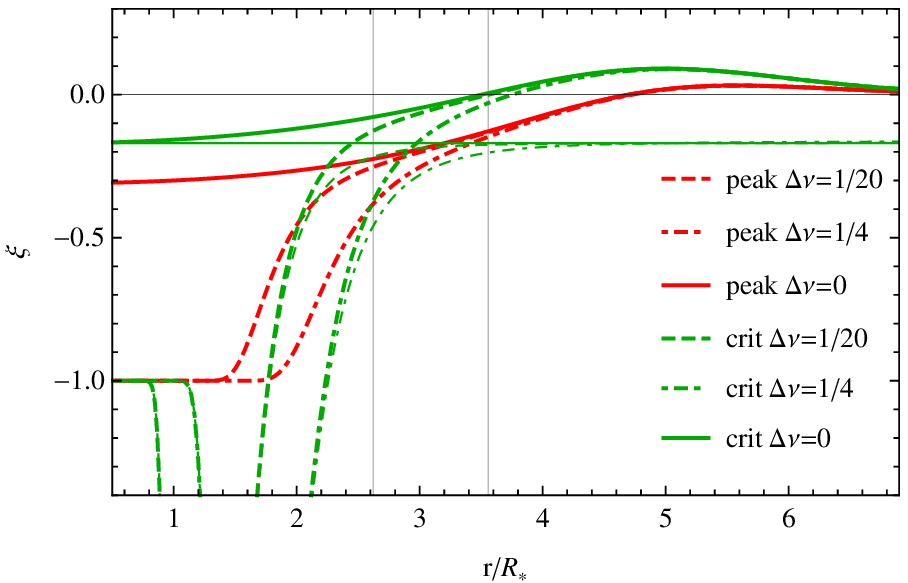}
\includegraphics[width=0.49\columnwidth]{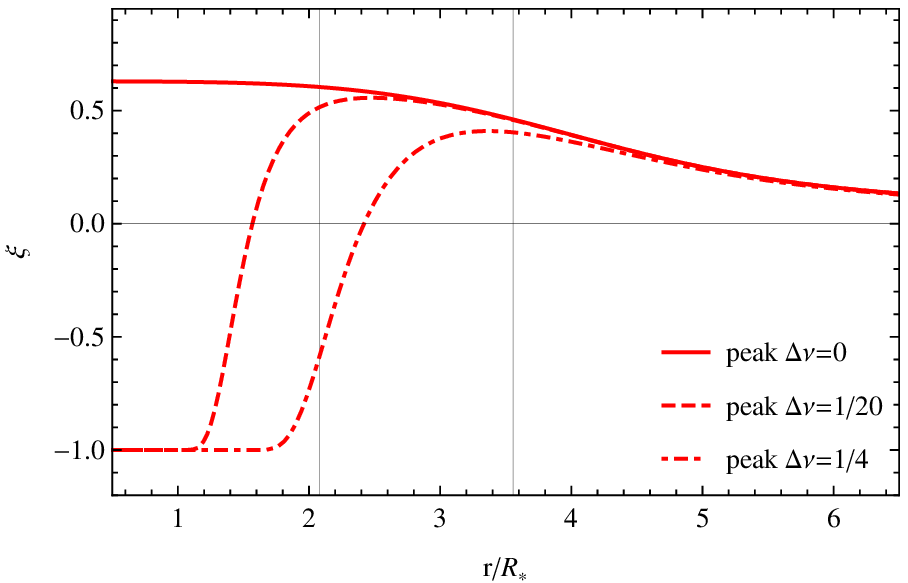}
\caption{\emph{Left panel: }Correlation function of peaks (red) and critical points (green) for $\bar \nu=3/2$ and for $\Delta \nu=0$ (solid), $\Delta \nu=1/10$ (dashed) and $\Delta \nu=1/4$ (dot-dashed). The scales, where the $\Delta \nu=0$ and $\Delta \nu\neq 0$ cases deviate are similar for peaks and critical points.
For the critical points we overplot the low-$r$ expansion equation~\eqref{eq:lowrdiff}. This expansion provides a good description of the low-$r$ behaviour up to 2-3 $R_\star$.
The vertical lines indicate the respective $r_{1\%}$ scales, and provide a useful estimate of the scale where the $\Delta \nu=0$ and $\Delta \nu\neq 0$ cases start to deviate in the full calculation. \emph{Right panel: }Same as left panel for a projected $\Lambda$CDM power spectrum with $R=2\; h^{-1}\text{Mpc}$.}
\label{fig:varyingnu2}
\end{figure}

%-------------------------------------------------------------------------
\label{sec:smallR}
So far, we have assumed that the peaks under consideration have exactly the same height. This
is clearly a very special case, since a realistic sample of haloes is likely to be made up by a
range of masses or smoothing scales and thus of different peak heights $\nu$.
As we have seen in Fig.~\ref{fig:1Dpeakcrit}, the exclusion region is often reduced to an
anti-correlation, in the sense that
$1+\xpk(r,\nu)$ does not reach zero at the origin. In order to ascertain whether this is a robust
feature, we have also computed the correlation function of 1D peaks and critical points with different 
heights $\nu_{1}=\nu-\Delta \nu/2$ and $\nu_{2}=\nu+\Delta \nu/2$, where $\Delta\nu>0$ is the 
height difference. For the critical points we have at small separations
\begin{equation}
\begin{split}
1+\xi_\text{crit}\left(\tilde r,\nu-\frac{\Delta \nu}{2},\nu+\frac{\Delta \nu}{2},n=0\right)=&\sqrt 3
\biggl[
\frac{\alpha_{8}(\nu,\Delta\nu) \Delta \nu ^2}{\tilde r^8 \left(\Delta \nu ^2-4 \nu^2\right)}
+
   \frac{\alpha_{6}(\nu,\Delta\nu) \Delta \nu ^2}{\tilde r^6 \left(\Delta \nu ^2-4 \nu
   ^2\right)}
   +
   \frac{\alpha_{4}(\nu,\Delta\nu)\Delta \nu ^2}{\tilde r^4 \left(\Delta \nu ^2-4 \nu
   ^2\right)}
   +
   \frac{\alpha_{2}(\nu,\Delta\nu) \Delta \nu ^2}{5 \tilde r^2 \left(\Delta \nu ^2-4 \nu
   ^2\right)} \\
   &+\frac{\alpha_{0}(\nu,\Delta\nu)}{ \left(\Delta \nu ^2-4 \nu^2\right)}
   \biggr]
   \exp\left[\frac{\nu ^2}{4}+\frac{7 \Delta \nu ^2}{80}-\frac{9 \Delta \nu ^2}{5 \tilde r^2}+\frac{27
   \Delta \nu ^2}{\tilde r^4}-\frac{324 \Delta \nu ^2}{\tilde r^6}\right]+{\cal O}(\tilde r)\,,
\end{split}
\label{eq:lowrdiff}
\end{equation}
where the functions $\alpha_{2n}(\nu,\Delta\nu)$ for $n$ between 0 and 4 are given in Appendix~\ref{sec:app}
 and
\begin{equation}
\alpha_{0}(\nu,\Delta\nu=0)=\left(\frac 1 8 +\frac{1}{3\nu^{2}}\right)\,.
\end{equation}
Comparing this expression to equation~\eqref{eq:lowrequal} above, we see that for unequal heights, inverse 
powers of the separation arise.
The $r^{-6}$ term in the exponential drives the correlation function to -1 on very small scales. We can 
estimate where the leading inverse power in the exponent causes a one percent correction to $1+\xi$ 
\begin{equation} 
r_{1\%}\approx \left(180 \Delta \nu\right)^{1/3}R_\star.
\end{equation}
Numerical results for peaks are shown in Fig.~\ref{fig:varyingnu} for different choices of 
$\Delta\nu$. While the correlation $1+\xpk(r,\nu_1, \nu_2)$ of peaks of different heights tends 
towards zero at small separation, it converges towards a finite non-zero value when the peaks
have exactly the same height. In Fig.~\ref{fig:varyingnu2}, we compare the exclusion of peaks and 
critical points to the
approximation of equation~\eqref{eq:lowrdiff}. The scales where the finite separation results deviate from the
equal height case are approximately the same for peaks and critical points and $r_{1\%}$ is a good 
indicator of this scale. We also compare the size of the exclusion region for peaks in a projected $\Lambda$CDM
density field with the corresponding $r_{1\%}$ and find equally good agreement.\footnote{For the peaks in a
projected $\Lambda$CDM power spectrum, we smooth the underlying 3D power spectrum with a Gaussian filter and
then consider the peaks along one of the coordinate axis (taken to be the $z$-axis without loss of generality). The effective one dimensional input power spectrum entering equation \eqref{eq:gammaij} is then
\be
P_{s,1\text{D}}(k_z)=\int \frac{\derd k_x \derd k_y}{(2\pi)^2}P_{3\text{D}}\left(\sqrt{k_x^2+k_y^2+k_z^2}\right)W^2\left(\sqrt{k_x^2+k_y^2+k_z^2}R\right)\; ,
\ee
where $W(k R)=e^{-k^2 R^2/2}$ is the Gaussian window.
}
Therefore, the behaviour of $\xpk$ in the limit $r\to 0$ strongly depends on the peak height
difference. This can be easily understood as follows: consider two peaks infinitesimally close to
each other. Very stringent constraints on the first and second derivatives of the density field
are thus required to bridge them. Clearly, the constraints will be more draconian the larger 
the height difference. Therefore, this configuration becomes increasingly unlikely with 
increasing $\Delta\nu>0$, and thus $1+\xpk(r,\nu_1,\nu_2)$ rapidly drops to zero as the 
separation decreases. The same behaviour is, of course, expected to hold for peaks with different 
smoothing scales and for integrals over bins in peak height. 
Note that the short distance behaviour of the peak correlation with fixed height does not follow
the exponential suppression $\exp(-R_\star ^2/r^2)$ found by \cite{lumsden/heavens/peacock:1989} for
peaks above a threshold.\footnote{{  As we have noted before, \cite{lumsden/heavens/peacock:1989} did not
go to sufficiently high orders in small scale separation to account for the cancellations of the elements of the
covariance matrix leading to the correct low-$r$ behaviour of the inverse covariance and determinant. This likely also invalidates their conclusions
on the leading behaviour of the peak correlation function above threshold.}}

%>>>>>>>>>>>>>>>>>>>>>>>>>>>>>>>>>>>>>>>>>>>>>>>>>>>>>>>>>>>>>>>>>>>>>>>>>>>>>>>>>>>>
%>>>>>>>>>>>>>>>>>>>>>>>>>>>>>>>>>>>>>>>>>>>>>>>>>>>>>>>>>>>>>>>>>>>>>>>>>>>>>>>>>>>>

%>>>>>>>>>>>>>>>>>>>>>>>>>>>>>>>>>>>>>>>>>>>>>>>>>>>>>>>>>>>>>>>>>>>>>>>>>>>>>>>>>>>>
%>>>>>>>>>>>>>>>>>>>>>>>>>>>>>>>>>>>>>>>>>>>>>>>>>>>>>>>>>>>>>>>>>>>>>>>>>>>>>>>>>>>>

%%%%%%%%%%%%%%%%%%%%%%%%%%%%%%%%%%%%%%%%%%%%%%%%%%%%%%%%%%%%%%%%%%%%%%%%%%
\section{Large scales: Perturbative bias expansion}
%%%%%%%%%%%%%%%%%%%%%%%%%%%%%%%%%%%%%%%%%%%%%%%%%%%%%%%%%%%%%%%%%%%%%%%%%%
\label{sec:bias}
Perturbative bias expansions have been widely used to predict clustering statistics of 
dark matter haloes and galaxies. However, no study so far has explored the convergence
properties of these series because of the highly non-linear (non-perturbative) effects
induced by small-scale exclusion. In this section, we wish to address this issue using
the clustering of 1D peaks as a proxy for the two-halo term. 

%-------------------------------------------------------------------------
\subsection{Methodology}
%-------------------------------------------------------------------------

As discussed in \cite{desjacques:2013}, the two-point correlation function of 3D peaks 
can be computed from a perturbative, local (peak) bias expansion in which the coefficients 
(bias parameters) are computed from a generalized peak-background split ansatz. 
This procedure is fairly general and it applies to any 'point' process of a Gaussian 
(and possibly non-Gaussian) random field. Therefore, it should certainly describe the 
two-point function of our 1D peak. In 1D, the perturbative bias expansion is constructed 
from three rotationally invariant quantities, i.e. the rescaled field $x$, $x_1^2$ and 
$x_{11}$ (in the notation of \cite{bardeen/bond/etal:1986}, $x=\nu$, $x_1=\eta/\sigma_{1}$ and $x_{11}=-x$).
Following equation~(\ref{eq:npk}), the ``localized'' number density of peaks of height $\nu$ 
then is
\begin{equation}
\npk(\boldsymbol{X}) = 
\frac{\sigma_2}{\sigma_1} |x_{11}| \delta_{\rm D}\!(x_1) \delta_{\rm D}\!(x-\nu)\Theta_{\rm H}(-x_{11})\,.
\end{equation}
The knowledge of $\npk(\nu)$ suffices to derive the bias parameters associated with 
this point process at all orders. Namely, the probability density for the variables
$\boldsymbol{X}=(x,x_{11},x_{1})$ is the product of a bivariate normal ${\cal N}(x,x_{11})$ 
with a chi-square distribution $\chi_1^2(x_1^2)$ with one degree of freedom. 
To construct the perturbative bias expansion, we proceed as in \cite{desjacques:2013}
and perturb the localized peak number density (the peak-background split). This
perturbation can be represented as a series in the appropriate orthogonal polynomials,
i.e. Hermite polynomials for ${\cal N}(x,x_{11})$ and generalized Laguerre polynomials for
$\chi_1^2(x_1^2)$. The perturbative bias expansion describing 1D peaks thus is
\begin{align}
\label{eq:PTbiasPK}
\dpk(r) &= \sum_{i,j,k} \frac{\sigma_0^i\sigma_2^j\sigma_1^{2k}}{i!j!}
\frac{\Gamma(1/2)}{\Gamma(k+1/2)} b_{ij}\chi_k H_{ij}\big(x(r),-x_{11}(r)\big)
L_k^{(-1/2)}\!\big(x_1(r)\big) \,, \\
&=\sigma_0 b_{10}x(r)-\sigma_2 b_{01}x_{11}(r)+\frac{1}{2}\sigma_0^2 b_{20}\big[x^2(r)-1\big]
-\sigma_0\sigma_2 b_{11}x(r) x_{11}(r)+\frac{1}{2}\sigma_2^2 b_{02}\big[x_{11}^2(r) -1\big]
+ \sigma_1^2\chi_1 \big[x_1^2(r)-1\big] +\dots\;, \nonumber 
\end{align}
where $\dpk$ is the mean field peak overabundance, and $r$ is the 1D comoving coordinate. 
The factor of unity in the quadratic terms remove the zero-lag contributions.
The bias parameters are the ensemble averages
\begin{align}
\label{eq:bias}
\sigma_0^i\sigma_2^j b_{ij} &= \frac{1}{\bnpk(\nu)}\int{\rm d}\boldsymbol{X}\, 
\npk(\boldsymbol{X})\, {\cal N}(\mathbf{X})\, H_{ij}(x,-x_{11})\,, \\
\label{eq:chi}
\sigma_1^{2k}\chi_k &= \frac{(-1)^k}{\bnpk(\nu)}\int{\rm d}\boldsymbol{X}\, \npk(\boldsymbol{X})\, 
{\cal N}(\boldsymbol{X})\, L_k^{(-1/2)}\!(x_1^2/2) \;.
\end{align} 
Here, $H_{ij}$ and $L_k^{(-1/2)}$ are bivariate Hermite and Laguerre polynomials, 
respectively so that
$\chi_k$ is simply $\chi_k=(-1/2)^k (2k-1)!!/(k!\sigma_1^{2k})$
and
$H_{ij}(x,-x_{11})={\cal N}^{-1}(x,x_{11})\left(-{\partial}_{x}\right)^{i}\left(\partial_{x_{11}}\right)^{j}{\cal N}(x,x_{11})$ 
has been commonly defined for the peak curvature $-x_{11}$. For instance,
\begin{align}
H_{10}(x,-x_{11})=&\frac{x+\gamma x_{11}}{1-\gamma^{2}}\\
H_{01}(x,-x_{11})=&-\frac{\gamma x+x_{11}}{1-\gamma^{2}}
.
\end{align}
The average peak number density entering equations~(\ref{eq:bias}-\ref{eq:chi}) is given by
\begin{align}
\bnpk(\nu) &= \frac{\sigma_2}{\sigma_1}\int{\rm d}\boldsymbol{X}\, {\cal N}(\boldsymbol{X})\,
|x_{11}| \delta_{\rm D}\!(x_1) \delta_{\rm D}\!(x-\nu)\Theta_{\rm H}(-x_{11}) \,,\\
&= \frac{\sigma_2}{\sigma_1}\frac{1}{\sqrt{2\pi}}\int{\rm d}x_{11}\,|x_{11}|\,
{\cal N}(\nu,x_{11}) \Theta_{\rm H}(-x_{11})\nonumber \,, \\
&= \frac{\sigma_2}{\sigma_1}\biggl\{\frac{1}{2}\gamma\nu
\left[1+{\rm Erf}\!\left(\frac{\gamma\nu}{\sqrt{2-2\gamma^2}}\right)\right]
+\sqrt{\frac{1-\gamma^2}{2\pi}}e^{-\frac{\gamma^2\nu^2}{2(1-\gamma^2)}}\biggr\}
\frac{e^{-\nu^2/2}}{2\pi} \nonumber \,, \\
&\equiv \frac{1}{2\pi R_\star}G(\gamma,\gamma\nu) e^{-\nu^2/2} \nonumber \;,
\end{align}
where, again, $R_\star=\sigma_1/\sigma_2$. 
%>>>>>>>>>>>>>>>>>>>>>>>>>>>>>>>>>>>>>>>>>>>>>>>>>>>>>>>>>>>>>>>>>>>>>>>>>>>>>>>>>>>>
\begin{figure}
\centering
\includegraphics[width=0.45\columnwidth]{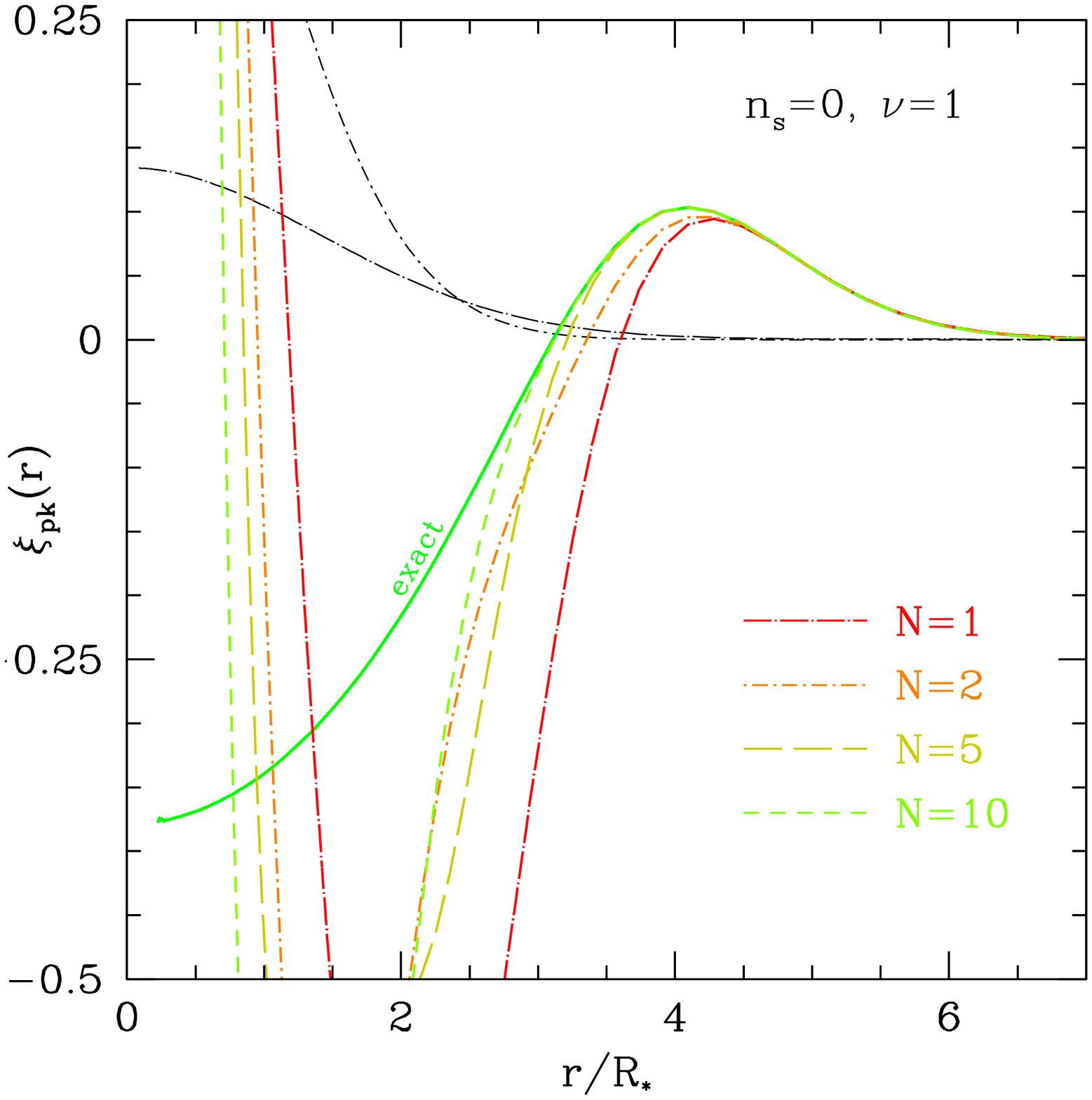}
\includegraphics[width=0.45\columnwidth]{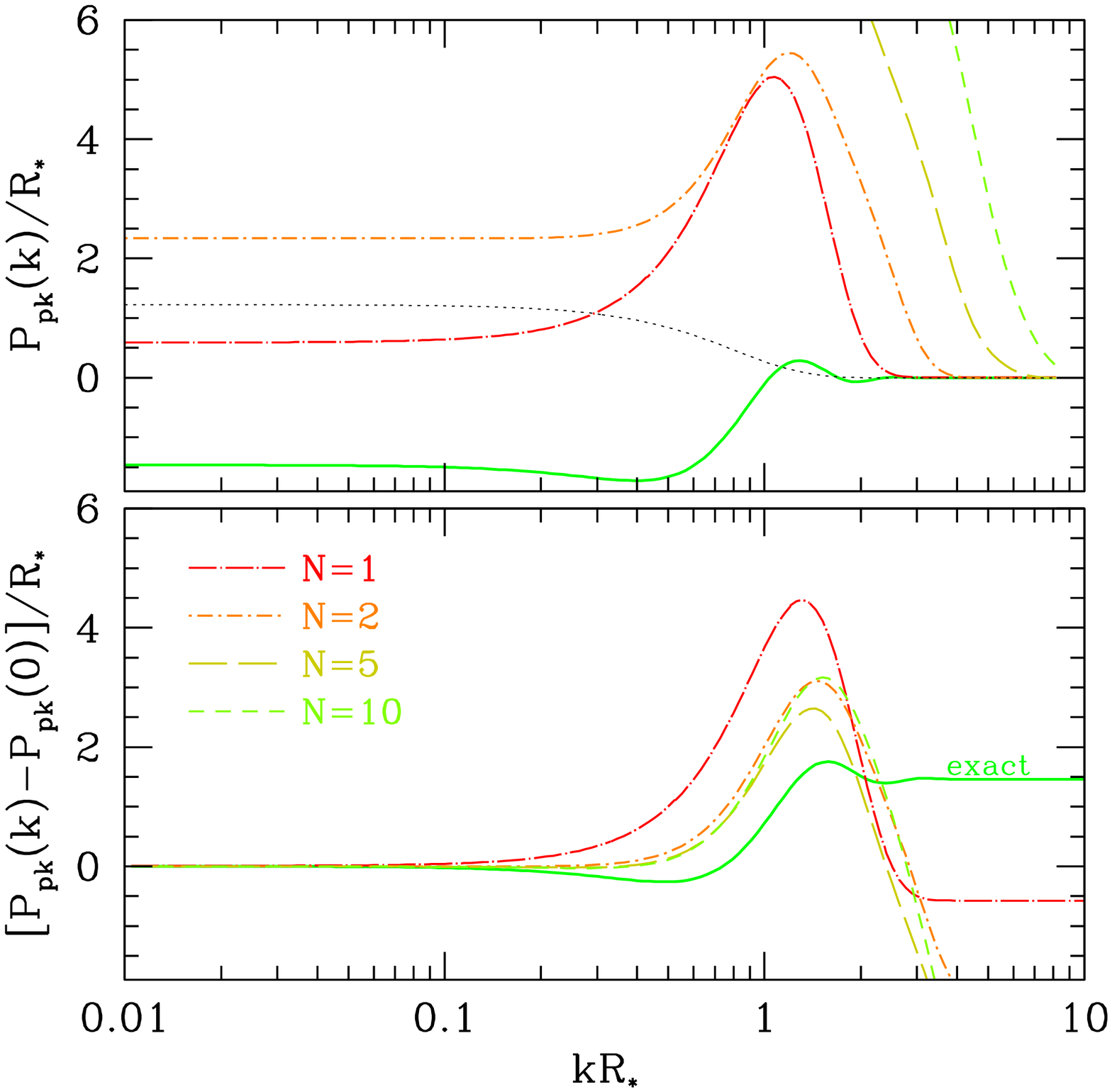}
\caption{{\it Left panel}: Comparison between the exact correlation function of 1D peaks and 
perturbative series obtained upon truncating the peak bias expansion equation~(\ref{eq:PTbiasPK}) 
at order $N=1$, 2, 5 and 10. For illustration, the black curves show linear and quadratic 
local bias approximations and perform significantly worse than the peak bias expansion. 
A white noise power spectrum with Gaussian filtering and a peak height $\nu=1$ were assumed. 
{\it Top right}: Exact peak power spectrum (solid) compared to various approximations 
obtained upon truncating the bias expansion at order $N=1$, 2, 5 and 10.  For comparison, the dotted line represents the linear bias approximation. {\it Bottom right}: same as top panel but the zero-lag value 
$P_\text{pk}(0)$ has been subtracted off from the power spectra. In all cases, a white noise 
power spectrum with Gaussian smoothing and a peak height of unity were assumed. }
\label{fig:ptbias}
\end{figure}
%>>>>>>>>>>>>>>>>>>>>>>>>>>>>>>>>>>>>>>>>>>>>>>>>>>>>>>>>>>>>>>>>>>>>>>>>>>>>>>>>>>>>
The peak bias expansion given by equation~(\ref{eq:PTbiasPK}) is to be compared with the standard linear bias approach which relates linearly the density of dark matter haloes to the dark matter density field
\begin{equation}
\delta_{h}(r) =\sigma_0 b\, x(r)\;,
\end{equation}
where $b\equiv b_{10}$ is the usual linear bias parameter.

In 1D, we could also use $x_1$ rather than $x_1^2$ as independent variable. $x_1$ is 
normally distributed, so that the joint probability distribution reads 
$P_1(x,x_1,x_{11})={\cal N}(x,x_{11}){\cal N}(x_1)$, and the bias parameters $\chi_k$, 
now associated to $x_1$, are
\begin{align}
\sigma_1^k\chi_k &= \frac{1}{\bnpk}\int{\rm d}\boldsymbol{X}\,{\cal N}(\boldsymbol{X})\,H_k(x_1) \\
& = (-1)^{k/2}\left(k-1\right)!! \qquad\mbox{if $k$ is even} \nonumber \;.
\end{align}
In both cases, we recover the same series expansion. 
Namely, on perturbing either ${\cal N}(x_1)$ or $\chi_1^2(x_1^2)$ with a long-wavelength 
perturbation $x_{1l}$ such that  $x_1\to x_1+x_{1l}$ and $x_1^2\to x_1^2+x_{1l}^2$, we find
\citep{desjacques:2013}
\begin{align}
\mbox{i)}&\qquad \sum_{n=0}^\infty \frac{1}{n!}(\sigma_1^n\chi_n)x_{1l}^n 
= \sum_{k=0}^\infty \frac{1}{(2k)!}(-1)^k(2k-1)!!\, x_{1l}^{2k}
= \sum_{k=0}^\infty\frac{(-1)^k}{2^k k!} x_{1l}^{2k}\,, \quad (\mbox{Hermite}) \nonumber \\
\displaybreak[1]
\mbox{ii)}&\qquad \sum_{k=0}^\infty \frac{\Gamma(\alpha+1)}{\Gamma(\alpha+k+1)}(\sigma_1^{2k}\chi_k) 
\left(\frac{x_{1l}^2}{2}\right)^k 
= \sum_{k=0}^\infty\frac{2^k}{(2k-1)!!}(-1/2)^k \frac{(2k-1)!!}{k!}
\left(\frac{x_{1l}^2}{2}\right)^k
= \sum_{k=0}^\infty\frac{(-1)^k}{2^k k!}x_{1l}^{2k} \quad (\mbox{Laguerre}) \nonumber \;.
\end{align}
In other words, we could equally work with the bias factors $\chi_k$ expressed either as 
Hermite or Laguerre.  We could also have followed \cite{gay/pichon/pogosyan:2012} and works 
with the variables $(z=(x+\gamma x_{11})/\sqrt{1-\gamma^2},x_{11})$ which have the advantage of 
being statistically independent. In what follows, we will stick to the variable $x_1^2$.

The 1D peak two-point correlation $\xpk(r)$ can be perturbatively computed from equation~(\ref{eq:PTbiasPK}).
In practice, one evaluates the ensemble average $\big\langle\npk({\boldsymbol r}_{x})\npk({\boldsymbol r}_{y})\big\rangle$, 
ensuring to discard all the terms involving zero-lag moments. At first order, this is
\begin{equation}
\xpk(r) = b_{10}^2 \xi_0(r) + 2 b_{10} b_{01} \xi_1(r) + b_{01}^2 \xi_2(r) \;, 
\end{equation}
where
\begin{equation}
\label{eq:xi-l}
\xi_l(r) = \frac{1}{\pi}\int_0^\infty\!\!{\rm d}k\,k^{2l} P_s(k)\times \bigg\lbrace
\begin{array}{ll} \cos(kr) & l ~\mbox{integer} \\ \sin(kr) & l ~\mbox{half integer} \end{array}\;.
\end{equation}
Note that $\xi_{i}(r)=\sigma_i^2\gamma_{ii}(r)$, where $\gamma_{ij}(r)$ is defined in 
equation~(\ref{eq:gammaij}).
For Gaussian initial conditions (which we assume throughout this paper), the $N$th order 
contribution to $\xpk(r)$ involves $N(N+1)/2$ distinct combinations of $x$ and $x_{11}$ 
correlators together with $N^2$ terms involving correlators of $x_{1}^2$. Therefore, the 
number of terms scales like $N^2$. Consequently, the 1D peak correlation up to $N$th order
involves ${\cal O}(N^3)$.
All these contributions can be expressed as a product of the 6 possible connected two-point 
correlators $\la x({\boldsymbol r}_{x})x({\boldsymbol r}_{y})\ra$, $\la x({\boldsymbol r}_{x})x_1({\boldsymbol r}_{y})\ra$ etc.  
However, the coefficients are product of $N$th order bias parameters and change from term 
to term. Therefore, the bias perturbative expansion will be of limited use unless the bias 
coefficients can be computed quickly. In practice, exploiting the recurrence among the 
orthogonal polynomials can help reducing the computational cost.

%-------------------------------------------------------------------------
\subsection{Convergence in real and Fourier space}
%-------------------------------------------------------------------------

Armed with these results, we can assess the convergence properties of the 1D peak perturbative 
bias expansion. For illustrative purposes, we will consider a power-law power spectrum 
$P_{s}(k)\propto k^n$ with $n=0$ and a peak height $\nu=1$.
In the left panel of Fig.~\ref{fig:ptbias}, we compare the exact result (solid green curve) 
with the $N=1$, 2, 5 and 10th order perturbative approximations (in colour). While the latter  
captures the excess correlation at a few $R_\star$ relatively well, the convergence to the 
exact result is fairly slow at shorter separations where exclusion effects become important.
In any case, the peak bias expansion equation~(\ref{eq:PTbiasPK}) performs significantly better than
a ``local bias'' approximation, in which only the dependence on $x(r)$ is retained in the 
perturbative series equation~(\ref{eq:PTbiasPK}). The first and second-order ``local bias''
approximations are shown as the black curves, and clearly furnish a poor fit to the exact result
for $r/R_\star\lesssim$ a few.

The upper right panel of Fig.~\ref{fig:ptbias} displays the resulting power spectra obtained 
by taking the Fourier transform of $\xpk(r)$ and its various perturbative approximations. For the
sake of comparison, the dotted black curve represents the first order local bias approximation. 
Small-scale exclusion translates into a white-noise contribution in the limit $k\to 0$ which 
makes the shot-noise non-Poissonian, in agreement with the findings of 
\cite{smith/scoccimarro/sheth:2007,baldauf:2013excl}. Note that such a $k^0$ tail also arises
in the clustering of thresholded regions \citep{beltran/durrer:2011}.
The magnitude of the white-noise correction changes with
the order $N$ of the approximation because it receives contributions from all orders. As shown
in the previous section, deriving an analytic expression valid throughout the exclusion zone is 
practically impossible and, therefore, there is no hope to obtain exact expressions for these 
deviations from Poisson noise. 
The bottom right panel of Fig.~\ref{fig:ptbias} displays the power spectra once the white 
noise correction $P_{\text{pk}}(k=0)$ (which generally depends on the order $N$ of the approximation) has 
been subtracted. The relatively slow convergence of the perturbative approximations to the 
1D peak power spectrum towards the exact result reflects the behaviour seen in configuration 
space. We expect that these considerations also hold for the convergence of perturbative bias 
expansions of actual dark matter haloes, even though the convergence rate may depend on the
dimensionality and the shape of the density power spectrum.

%%%%%%%%%%%%%%%%%%%%%%%%%%%%%%%%%%%%%%%%%%%%%%%%%%%%%%%%%%%%%%%%%%%%%%%%%%
\section{Dynamics of the Peaks}
%%%%%%%%%%%%%%%%%%%%%%%%%%%%%%%%%%%%%%%%%%%%%%%%%%%%%%%%%%%%%%%%%%%%%%%%%%
\label{sec:zeldo}

So far we have been concerned with peaks in Lagrangian space, but haloes are observed in Eulerian space, i.e. after gravitational evolution.
{  In this Section we will identify peaks in Lagrangian space and move them according to their Zel'dovich displacement. The discrete tracer set generated in this way should be in close correspondence with the peaks identified in the evolved non-Gaussian field, or haloes in a more realistic setting.
This is the approximation that was made in the 3-dimensional calculation of \cite{desjacques/crocce/etal:2010}.}
For that purpose, let us use the Lagrangian displacement $\Psi$ to relate the Lagrangian peak position $q$ to its final Eulerian position $r$
\be
r_\text{pk}=q_\text{pk}+D_{\!+} \Psi_\text{pk}(q_\text{pk})\; ,
\ee
where $D_{\!+}$ is the linear growth factor and the displacement satisfies the equation of motion
\be
\Psi''+\mathcal{H} \Psi'=-\nabla_{r} \phi\; ,
\ee
where $\phi$ is the gravitational potential.
The linear solution to this equation in three dimensions is known as the Zel'dovich approximation and describes ballistic evolution according to the initial velocity. 
We will assume that the peak displacement is related to the mean displacement of the peak patch, i.e., by $\Psi(k)=-\ii k/k^2 \delta(k) W_R(k)$, where $W_R(k)$ is the smoothing window. {  Due to the presence of the smoothing scale we expect the Zel'dovich approximation to work better for the finite size peaks than for the matter field itself, which undergoes shell crossing.}
Conveniently, in one dimension, the Zel'dovich solution is the exact solution \citep{buchert:1989,yoshisato/matsubara/morikawa:1998,mcquinn:2015}.\footnote{From $[1+\delta(r)]\derd r=\derd q$ we have $\delta(r)=1/J-1$ with $J=|1+\nabla_q \Psi |$. Then we can take the Eulerian divergence of the equation of motion
\be
\nabla_r \Bigl[ \Psi''+\mathcal{H}\Psi'\Bigr]=\frac{3}{2}\mathcal{H}^2 \delta
\ee
Rewriting the Eulerian as a Lagrangian derivative $\nabla_r=(1+\nabla_q \Psi)^{-1}\nabla_q$ we finally have
\be
\nabla_q \Bigl[ \Psi''+\mathcal{H}\Psi'\Bigr]=\frac{3}{2}\mathcal{H}^2 \nabla_q \Psi
\ee
which is a linear equation for $\Psi$ that can be solved exactly by $\Psi=D_{\!+} \Psi_0$ before shell crossing, where $D_{\!+}$ is the linear growth.
}

%-------------------------------------------------------------------------
\subsection{Zel'dovich displacement of peaks}
%-------------------------------------------------------------------------
We can now write the Eulerian halo/peak overdensity as a sum over Eulerian peak positions, which are in turn related to their respective Lagrangian proto-halo positions
\begin{align}
1+\delta_\text{pk}(r)=&\frac{1}{\bar n_\text{pk}}\sum_\text{pk} \ddir\left(r- r_\text{pk}\right)=\int \derd q' \ddir\left[ r- q'- D_{\!+}\Psi( q')\right]\sum_\text{pk}\ddir( q'- q_\text{pk})\,,\\
=&\frac{\sigma_2}{\sigma_1}\int \derd q' \int \frac{\derd k}{2\pi}\eh{\ii  k( r- q')}|x_{11}(q')|\ddir\bigl[x_1( q')\bigr]\Theta_{\rm H}\bigl[-x_{11}(q')\bigr]\eh{-\ii  kD_{\!+}  \Psi( q')}\,,
\end{align}
where we expanded the Dirac function as plane waves.
The correlation of evolved peaks is therefore given by
\begin{align}
\xi(r)=\la\delta_\text{pk}(0)\delta_\text{pk}(r)\ra=&\frac{1}{\bar n_\text{pk}^2}\frac{\sigma_2^2}{\sigma_1^2}\int \derd Q  \int \frac{\derd k}{2\pi}  \eh{\ii k (Q-r)}\Bigl\langle \eh{-\ii k D_{\!+} (\Psi_1-\Psi_2)} \nonumber\\
&\times\  |x_{11}(q_1)| |y_{11}(q_2)| \ddir\bigl[x_1( q_1)\bigr]\ddir\bigl[y_1( q_2)\bigr]\Theta_{\rm H}\bigl[-x_{11}(q_1)\bigr]\Theta_{\rm H}\bigl[-y_{11}(q_2)\bigr]\Bigr\rangle-1\; ,
\end{align}
where $Q=q_2-q_1$ is the Lagrangian separation of the peaks and $\Psi_1$ and $\Psi_2$ are the displacements at the respective positions.
The displacement field is an additional stochastic variable that needs to be averaged over under the peak constraint. For this purpose, we need to append the covariance matrix by components that describe the auto-covariance of the displacement and the cross-covariance between the displacement on the one hand and the density field and its derivatives on the other hand. 
Splitting the state vector into $\boldsymbol \Phi=(\boldsymbol X,\boldsymbol\Psi)$ where in this section we redefine $\boldsymbol X$ as $\boldsymbol X=(\sigma_0 x,0,\sigma_2 x_{11},\sigma_0 y,0,\sigma_2 y_{11})$ and 
$\boldsymbol\Psi=(\Psi_1,\Psi_2)$, the covariance matrix (see Appendix~\ref{app:cov} for an explicit expression) takes the following schematic form 
\be
\mathbf{C}=
\begin{pmatrix}
\la \boldsymbol{ X}\!\cdot\! \boldsymbol X^{\rm T}\ra & \la \boldsymbol X \!\cdot\! \boldsymbol\Psi^{\rm T}\ra\\
\la \boldsymbol\Psi\!\cdot\!\boldsymbol X^{\rm T}\ra & \la \boldsymbol\Psi\!\cdot\!\boldsymbol\Psi^{\rm T }\ra
\end{pmatrix}
=
\begin{pmatrix}
\mathbf{C}_{ {X}} & \mathbf{C}_{{X\Psi}}\\
\mathbf{C}_{{X\Psi}}^{\rm T}&\mathbf{C}_{ {\Psi}}
\end{pmatrix}
=
\begin{pmatrix}
\mathbf{\Omega}_{ {X}} & \mathbf{\Omega}_{{X\Psi}}\\
\mathbf{\Omega}_{{X\Psi}}^{\rm T}&\mathbf{\Omega}_{{\Psi}}
\end{pmatrix}^{-1}
=\mathbf\Omega^{-1}\,,
\ee
where $\mathbf\Omega=\mathbf C^{-1}$ is the precision matrix.
The determinant of the covariance matrix can be decomposed into $\det \mathbf{C}=
\det \mathbf{C}_{ {X}}/
\det\mathbf{\Omega}_{ {\Psi}}$ 
with $\mathbf{\Omega}_{ {\Psi}}=
(\mathbf{C}_{ {\Psi}}-\mathbf{C}_{{\!X\!\Psi}}^\textrm{T}\cdot\mathbf{C}_{ {X}}^{-1}\cdot\mathbf{C}_{{\!X\!\Psi}})^{-1}$, which will prove useful in the rest of the calculation.
Defining $\ii k(\Psi_1-\Psi_2)=\ii k \,\boldsymbol u^{\rm T}\!\cdot\!\boldsymbol\Psi$ with $\boldsymbol u=(1,-1)$, one can rewrite the argument of the exponential
arising from the PDF and the shift as
\begin{align}
\boldsymbol \Phi^{\rm T}\!\cdot\! \mathbf{C}^{-1}\!\!\cdot\!\boldsymbol \Phi\!+\!2\ii k D_{\!+} \,\boldsymbol u^{\rm T} \!\!\cdot\!\boldsymbol\Psi=&\boldsymbol X^{\rm T}\!\!\cdot\! \mathbf{\Omega}_{ {X}}\!\!\cdot\! \boldsymbol X\!+\!2\boldsymbol X^{\rm T}\!\!\cdot\! \mathbf{\Omega}_{{\!X\!\Psi}}\!\!\cdot\!\boldsymbol\Psi \!+\!\boldsymbol\Psi^{\rm T} \!\!\cdot\!\mathbf{\Omega}_{ {\Psi}}\!\!\cdot\!\boldsymbol\Psi\!+\!2\ii D_{\!+} k \,\boldsymbol u^{\rm T}\!\!\cdot\! \boldsymbol\Psi\nonumber\\
=&\boldsymbol X^{\rm T}\!\!\cdot\! \mathbf{\Omega}_{ {X}} \!\!\cdot\!\boldsymbol X \!+\!2\boldsymbol X^{\rm T} \!\!\cdot\!\mathbf{\Omega}_{{\!X\!\Psi}}\!\!\cdot\! \boldsymbol\Psi\!+\!2 \boldsymbol \mu^{\rm T}\!\!\cdot\! \mathbf{\Omega}_{ {\Psi}}\!\!\cdot\!\boldsymbol\Psi\!+\!2 \ii D_{\!+} k \,\boldsymbol u^{\rm T}\! \!\cdot\!\boldsymbol\Psi\!-\!\boldsymbol \mu^{\rm T} \!\!\cdot\!\mathbf{\Omega}_{ {\Psi}}\!\!\cdot\! \boldsymbol \mu\!+\!(\boldsymbol\Psi-\boldsymbol \mu)^{\rm T} \!\!\cdot\!\mathbf{\Omega}_{ {\Psi}} \!\!\cdot\!(\boldsymbol\Psi-\boldsymbol \mu) ,
\end{align}
for any vector $\boldsymbol\mu$. Let us now 
complete the square to isolate the displacement part of the PDF and therefore choose
$\boldsymbol \mu^{\rm T}=-(\boldsymbol X^{\rm T}\!\!\cdot\! \mathbf{\Omega}_{{\!X\!\Psi}}+ \ii D_{\!+} k \,\boldsymbol u^{\rm T})\!\cdot\!\mathbf{\Omega}_{ {\Psi}}^{-1}$. In that case, we get
\be
\boldsymbol \Phi^{\rm T}\!\!\cdot\! \mathbf{C}^{-1}\!\!\cdot\!\boldsymbol \Phi+2\ii k D_{\!+}\,\boldsymbol u^{\rm T} \!\!\cdot\!\boldsymbol\Psi=\boldsymbol X^{\rm T}\!\!\cdot\! \mathbf{C}_{ {X}} ^{-1}\!\!\cdot\! \boldsymbol X+D_{\!+}^2 k^2 \,\boldsymbol u^{\rm T} \!\!\cdot\!\mathbf{\Omega}_{ {\Psi}}^{-1}\!\!\cdot\!\boldsymbol u+2\ii D_{\!+} k \boldsymbol X^{\rm T} \!\!\cdot\!\mathbf{C}_{ {X}} ^{-1} \!\!\cdot\! \mathbf{C}_{{\!X\!\Psi}} \!\!\cdot\!\boldsymbol u+(\boldsymbol\Psi-\boldsymbol \mu)^{\rm T}\!\!\cdot\! \mathbf{\Omega}_{ {\Psi}}\!\!\cdot\! (\boldsymbol\Psi-\boldsymbol \mu).
\ee
The displacement vector $\boldsymbol\Psi$ can be integrated out, yielding unity and leaves us with
\begin{align}
\la\delta_\text{pk}(0)\delta_\text{pk}(r)\ra=&\frac{1}{\bar n_\text{pk}^2}\int \derd Q  \int \frac{\derd k}{2\pi} \int \derd \boldsymbol X  \eh{\ii k (Q-r)}\nonumber\\
&\frac{w(\boldsymbol X)}{\sqrt{(2\pi)^6 \det \mathbf{C}_{ {X}} }}\eh{-\frac12 \boldsymbol X^{\rm T}\!\!\cdot\! \mathbf{C}_{ {X}} ^{-1} \!\!\cdot\! \boldsymbol X-\frac12 D_{\!+}^2 k^2 \,\boldsymbol u^{\rm T} \!\!\cdot\! \mathbf{\Omega}_{ {\Psi}}^{-1}\!\!\cdot\! \,\boldsymbol u-\ii D_{\!+} k \boldsymbol X^{\rm T} \!\!\cdot\! \mathbf{C}_{ {X}} ^{-1} \!\!\cdot\! \mathbf{C}_{{\!X\!\Psi}} \!\!\cdot\! \, \boldsymbol u}
-1\, ,
\label{eq:movedpeak1d}
\end{align}
where the joint peak condition is defined as
\begin{equation}
w(\boldsymbol X)=|x_{11}(q_1)| |y_{11}(q_2)| \ddir\bigl[x_1( q_1)\bigr]\ddir\bigl[y_1( q_2)\bigr]\Theta_{\rm H}\bigl[-x_{11}(q_1)\bigr]\Theta_{\rm H}\bigl[-y_{11}(q_2)\bigr].
\end{equation}
Let us try to understand the linearized version of this result.
Fourier transforming in the Eulerian separation $r$ and expanding this expression for small correlations (i.e. $Q\to \infty$), we get to first order
\begin{align}
P_\text{pk}(k)=&\frac{1}{\bar n_\text{pk}^2}\int \derd Q \int \derd \boldsymbol X w(\boldsymbol X)\Bigl (-\frac12 \boldsymbol X^{\rm T}\!\!\cdot\! \mathbf{C}_{ {X}} ^{-1} \!\!\cdot\! \boldsymbol X-\frac12 D_{\!+}^2 k^2 \,\boldsymbol u^{\rm T} \!\!\cdot\! \mathbf{\Omega}_{ {\Psi}}^{-1}\!\!\cdot\! \,\boldsymbol u-\ii D_{\!+} k \boldsymbol X^{\rm T} \!\!\cdot\! \mathbf{C}_{ {X}} ^{-1} \!\!\cdot\! \mathbf{C}_{{\!X\!\Psi}} \!\!\cdot\! \, \boldsymbol u\Bigr)_{\mathcal{O}(\xi^1)}\nonumber\\
&\times \eh{\ii k Q}\eh{-D_{\!+}^2k^2 \sigma_{v,\text{pk}}^2}\eh{-\frac{1}{2}\frac{(x_{11}+\gamma x)^2}{1-\gamma^2}-\frac{1}{2}x^2}\eh{-\frac{1}{2}\frac{(y_{11}+\gamma y)^2}{1-\gamma^2}-\frac{1}{2}y^2}\; .
\end{align}
The expansions up to first order in correlations of the three constituent terms of the exponent are given by
\begin{eqnarray}
-\frac 1 2 \Bigl (\boldsymbol X^{\rm T}\!\!\cdot\! \mathbf{C}_{ {X}} ^{-1} \!\!\cdot\! \boldsymbol X\Bigr)_{\mathcal{O}(\xi^1)}&=&H_{10}(x,-x_{11})H_{10}(y,-y_{11})\frac{\xi_0}{\sigma_{0}^{2}}+H_{01}(x,-x_{11})H_{01}(y,-y_{11})\frac{\xi_2}{\sigma_{2}^{2}}
\nonumber\\
&&+\left[H_{10}(x,-x_{11})H_{01}(y,-y_{11})+H_{10}(y,-y_{11})H_{01}(x,-x_{11})\right]\frac{\xi_1}{\sigma_{0}\sigma_{2}}\; ,\\
-\frac 1 2 \Bigl (\boldsymbol u^{\textrm{T}}  \!\!\cdot\! \mathbf{\Omega}_{ {\Psi}}^{-1} \!\!\cdot\! \,\boldsymbol u\Bigr)_{\mathcal{O}(\xi^1)}&=& 
\xi_{-1}-2\xi_{0}\frac{\sigma_{0}^2}{\sigma_1^2}+\xi_1 \frac{\sigma_{0}^4}{\sigma_1^4} \,,\\
\Bigl (\boldsymbol X^\textrm{T}  \!\!\cdot\!  \mathbf{C}_{ {X}} ^{-1}  \!\!\cdot\!  \mathbf{C}_{{\!X\!\Psi}} \!\!\cdot\!  \,\boldsymbol u
\Bigr)_{\mathcal{O}(\xi^1)}&= &\frac{1}{\sigma_{0}}\bigl[H_{10}(x,-x_{11})+H_{10}(y,-y_{11})\bigr]\left(\xi_{-1/2}-\frac{\sigma_0^2}{\sigma_1^2}\xi_{1/2}\right)\,\nonumber\\
&&+\frac{1}{\sigma_{2}}\bigl[H_{01}(x,-x_{11})+H_{01}(y,-y_{11})\bigr]\left(\xi_{1/2}-\frac{\sigma_0^2}{\sigma_1^2}\xi_{3/2}\right)\, ,
\end{eqnarray}
where $H_{ij}$ is defined in Section~\ref{sec:bias}
and the peak velocity/displacement dispersion is given by
\be
\sigma_{v,\text{pk}}^2=\sigma_{-1}^2-\frac{\sigma_{0}^4}{\sigma_1^2}\,.
\label{eq:rmsveldisp}
\ee
Note that the correlation functions $\xi_i$ are defined in equation~\eqref{eq:xi-l} and the $\sigma_i$ in equation~\eqref{eq:sigma}.
Performing the integration over $Q$ amounts to Fourier transforming the correlation functions in the above expressions. Performing this Fourier transform and averaging over the  peak curvature variables $x_{11}$ and $y_{11}$, we have
\begin{align}
P_\text{pk}(k)\approx &\biggl[b_{10}(\nu_1)+b_{01}(\nu_1)k^2+D_{\!+}\left(1-\frac{\sigma_0^2}{\sigma_1^2}k^2\right) \biggr]\biggl[b_{10}(\nu_2)+b_{01}(\nu_2)k^2+D_{\!+}\left(1-\frac{\sigma_0^2}{\sigma_1^2}k^2\right)\biggr]P_s(k)\eh{-D_{\!+}^2k^2 \sigma_{v,\text{pk}}^2}\;.
\end{align}
The propagator term $\eh{-D_{\!+}^2k^2 \sigma_{v,\text{pk}}^2}$ arising here is an artifact of our na\"{\i}ve expansion, its long wavelength contributions should cancel for equal-time correlators. Furthermore, the above expression does not contain the constant stochasticity corrections that are arising from small-scale effects -- e.g exclusion -- in the correlation function, where $\xi_i\approx \sigma_i^2$ and the above expansion fails.

At low wavenumbers and defining again $b\equiv b_{10}$, the peak density power spectrum simplifies to
\begin{align}
P_\text{pk}(k)\approx &\bigl[b(\nu_1)+D_{\!+} \bigr]\bigl[b(\nu_2)+D_{\!+}\bigr]P_s(k),
\end{align}
meaning that the large-scale bias is given by $b+D_{\!+}$ as one would have expected from \cite{mo/white:1996}. In general however, the linear Lagrangian peak bias $b_{10}+b_{01} k^2$ decays to the velocity bias $1-\sigma_0^2/\sigma_1^2 k^2$ as discussed thoroughly in \cite{desjacques/sheth:2010,desjacques/crocce/etal:2010,baldauf/desjacques/seljak:2014}. In configuration space this statistical velocity bias appears as $\psi_\text{pk}=\psi-\sigma_0^2/\sigma_{1} x_1$ \citep{desjacques:2008ba}.

%>>>>>>>>>>>>>>>>>>>>>>>>>>>>>>>>>>>>>>>>>>>>>>>>>>>>>>>>>>>>>>>>>>>>>>>>>>>>>>>>>>>>
\begin{figure}
\centering
\includegraphics[width=0.49\textwidth]{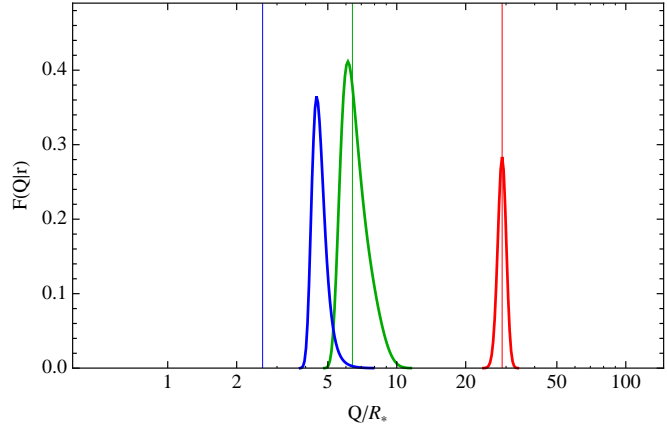}
\caption{Contribution to the integral over Lagrangian separations in equation~\eqref{eq:lagconv} for $D_+=7$. The vertical lines indicate the Eulerian separation $r$ and the corresponding curves indicate the support of the integral. We see that for large scales, the evolution merely corresponds to a convolution with a Gaussian, whereas for small scales there is a an offset in the support to larger scales and the convolution kernel is skewed.}
\label{fig:displcontrib}
\end{figure}
%>>>>>>>>>>>>>>>>>>>>>>>>>>>>>>>>>>>>>>>>>>>>>>>>>>>>>>>>>>>>>>>>>>>>>>>>>>>>>>>>>>>>

Having shown the consistency with previous 3-dimensional studies, let us come back to the full expression given by equation~\eqref{eq:movedpeak1d}. Together with the prefactor $\eh{\ii k (r-Q)}$, this is a Gaussian integral in $k$, which is readily performed and yields
\begin{equation}
\la\delta_\text{pk}(0)\delta_\text{pk}(r)\ra=\int \derd Q\ F(Q|r)-1\,,
\label{eq:lagconv}
\end{equation}
where
\begin{equation}
F(Q|r)=\frac{1}{\bar n_\text{pk}^2} \! \int\! \derd \boldsymbol X\! \frac{w(\boldsymbol X)}{\sqrt{(2\pi)^6 \det \mathbf{C}_{ {X}} }}\eh{-\frac12 \boldsymbol X^{\rm T}  \!\!\cdot\! \mathbf{C}_{ {X}} ^{-1}  \!\!\cdot\! \boldsymbol X}
 \frac{1}{\sqrt{2\pi D_{\!+}^2 \boldsymbol u^{\rm T}  \!\!\cdot\! \mathbf{\Omega}_{ {\Psi}}^{-1} \!\!\cdot\! \,\boldsymbol u }}\eh{-\frac{1}{2}\frac{(Q-r-D_{\!+} \boldsymbol X^{\rm T} \!\!\cdot\!  \mathbf{C}_{ {X}} ^{-1}  \!\!\cdot\! \mathbf{C}_{{\!X\!\Psi}} \!\!\cdot\! \,\boldsymbol u)^2}{D_{\!+}^2 \boldsymbol u^{\rm T}\!\!\cdot\!\mathbf{\Omega}_{ {\Psi}}^{-1}\!\!\cdot\!\boldsymbol u}},
\end{equation}
which is a convolution between the Eulerian and Lagrangian distances $r$ and $Q$. For a fixed Eulerian distance, the integral is approximately Gaussian and peaks at $Q_p=r+D_{\!+} \boldsymbol X^{\rm T} \!\!\cdot\!  \mathbf{C}_{ {X}} ^{-1}  \!\!\cdot\! \mathbf{C}_{{\!X\!\Psi}}  \!\!\cdot\! \,\boldsymbol u$ (as illustrated in Fig.~\ref{fig:displcontrib}).
As the growth factor $D_{\!+}$ goes to zero, the last part of the above integral becomes a Dirac delta function for $Q-r$ and we recover the Lagrangian expression. The shift term $D_{\!+} \boldsymbol X^{\rm T} \!\!\cdot\!  \mathbf{C}_{ {X}} ^{-1} \!\!\cdot\!  \mathbf{C}_{{\!X\!\Psi}}  \!\cdot\! \,\boldsymbol u$ 
corresponds by construction to the most likely  difference of Zel'dovich displacements, $\boldsymbol \Psi^{\rm T}\cdot \boldsymbol u$, (in the Wiener filtering sense) for a given $\boldsymbol {X}$; it   
is negligible on large scales, but can lead to significant corrections on small scales so that the convolution by these conditional displacements  actually fill the $\xi=-1$ region.

\begin{figure}
\centering
\includegraphics[width=0.49\textwidth]{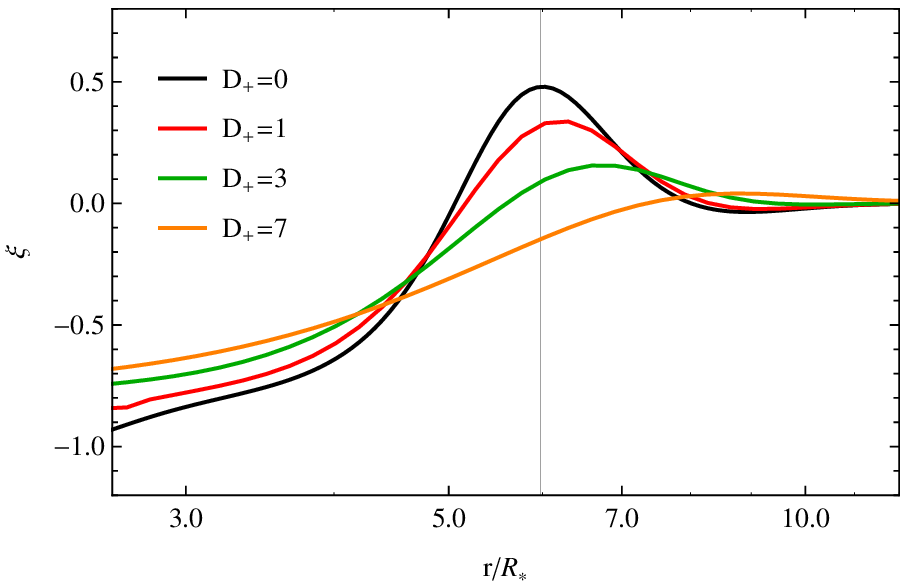}
\includegraphics[width=0.49\textwidth]{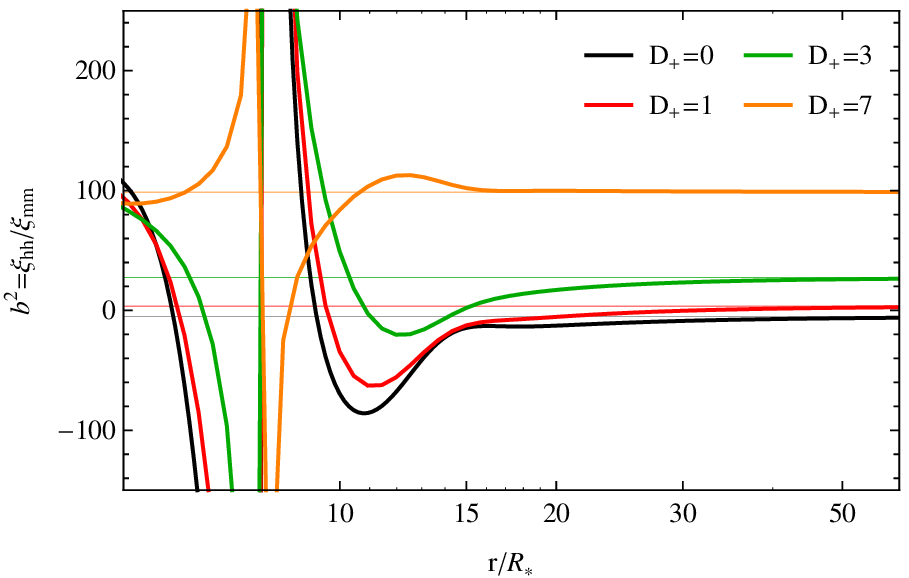}
\caption{\emph{Left panel: }Small-scale correlation function of initial and evolved peaks in a $n=3$ power law density field. We have chosen peaks of height $\bar \nu=6/5$ and $\Delta \nu=3/5$ to highlight the behaviour in presence of exclusion. The linear growth factor increases from the initial conditions $D_{\!+}=0$ to some arbitrary final time $D_{\!+}=7$. \emph{Right panel: }Linear bias of the peak correlation functions with respect to the linear matter correlation function. We clearly see that the bias asymptotes to the expected $b_{10}+D_{\!+}$ behaviour for large separations indicated by the horizontal lines.}
\label{fig:evolvedpeaks}
\end{figure}

In order to compute the effect of the peak displacements, we have to resort to one dimensional power spectra that grow steeper than $k^{ 1}$ for low wavenumbers since otherwise the velocity correlators diverge in the infrared ($k^{n-2}$ is not integrable). In particular, we need a power spectrum that has a slope $n>{  1}$ and we will choose $n=3$ for definiteness.
We show the evolved correlation function and bias in Fig.~\ref{fig:evolvedpeaks}. 
Non-linearities tend to wash out the exclusion and decrease the small-scale clustering of peaks.

\begin{figure}
\centering
\includegraphics[width=0.49\textwidth]{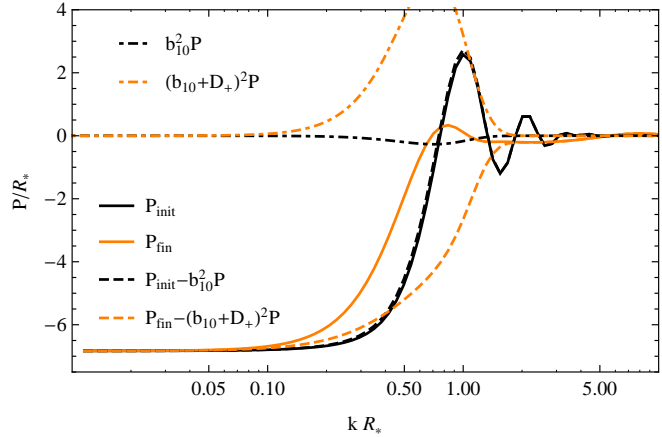}
\caption{Evolution of the peak-peak power spectrum and its noise component $P_\text{pk}-b^2 P_{s}$ from the initial conditions to the final configuration at $D_{\!+}=7$. While the amplitude of the stochasticity correction remains constant on large scales, the infall causes the transition to zero to happen on smaller scales/larger wavenumbers.}
\label{fig:evolvedpeaksp}
\end{figure}

In \cite{baldauf:2013excl} it has been shown in simulations that the stochasticity amplitude on large scales is the same in Lagrangian and Eulerian space. On small scales the corrections to the fiducial $1/\bar n$ stochasticity have to vanish, and they do so for haloes in Eulerian and proto-haloes in Lagrangian space. The only change in the behaviour is that the transition happens at higher wavenumbers in Eulerian space. We would like to explore to what extend our 1D peak model can reproduce this behaviour. In Fig.~\ref{fig:evolvedpeaksp} we show the power spectrum of peaks in Lagrangian space and their stochasticity estimated as $P_\text{pk}-b^2 P_{s}$. We see that the large-scale amplitude is the same for initial and evolved haloes in agreement with numerical studies. After correction for the large-scale bias in Eulerian space, we also recover that the transition between the non-zero and zero stochasticity correction regimes is pushed to higher wavenumbers.

%>>>>>>>>>>>>>>>>>>>>>>>>>>>>>>>>>>>>>>>>>>>>>>>>>>>>>>>>>>>>>>>>>>>>>>>>>>>>>>>>>>>>
%>>>>>>>>>>>>>>>>>>>>>>>>>>>>>>>>>>>>>>>>>>>>>>>>>>>>>>>>>>>>>>>>>>>>>>>>>>>>>>>>>>>>

%>>>>>>>>>>>>>>>>>>>>>>>>>>>>>>>>>>>>>>>>>>>>>>>>>>>>>>>>>>>>>>>>>>>>>>>>>>>>>>>>>>>>
%>>>>>>>>>>>>>>>>>>>>>>>>>>>>>>>>>>>>>>>>>>>>>>>>>>>>>>>>>>>>>>>>>>>>>>>>>>>>>>>>>>>>

%-------------------------------------------------------------------------
\subsection{Streaming Motions of Peaks}
\begin{figure}
\centering
\includegraphics[width=0.49\textwidth]{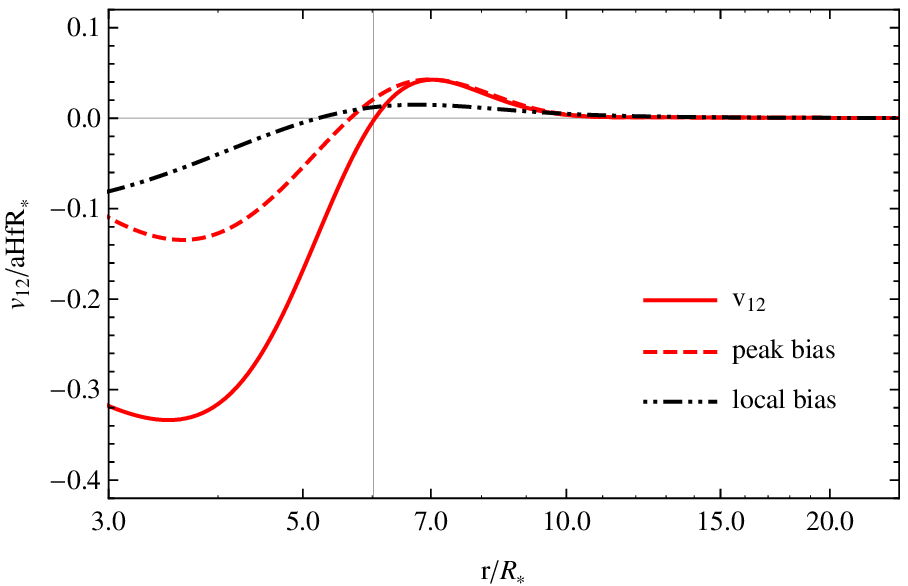}
\includegraphics[width=0.49\textwidth]{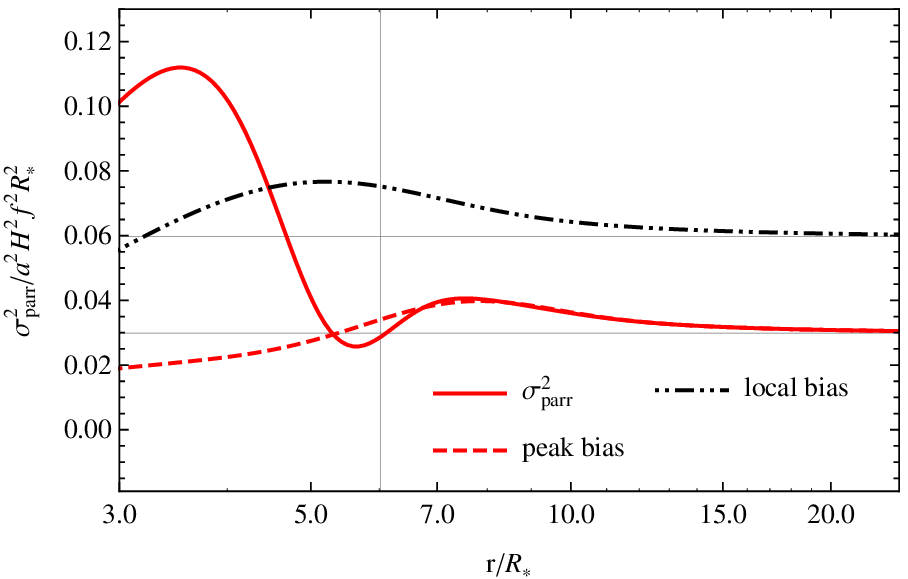}
\caption{\emph{Left panel: }Mean relative velocity $v_{12}$ of 1D peaks (red solid). The peak velocities clearly deviate from the velocities of random dark matter particles (black dot-dashed) on small scales. On large scales this deviation can be described by a linear velocity bias (red dashed), which however fails to describe the non-perturbative behaviour in the exclusion regime.
We see that below $r=6 R_\star$ peaks are moving towards each other and away from each other for larger separations. The vertical gray line indicates the peak of the correlation function in Fig.~\ref{fig:evolvedpeaks}. \emph{Right panel: }Velocity dispersion of the peaks in 1D. Even the large separation limit of the peak velocity dispersion deviates from the underlying dark matter, i.e. local bias (black dot-dashed), due to the correction in the peak displacement dispersion Eq.~\eqref{eq:rmsveldisp}. This difference and the onset of the scale-dependence is described by the linear scale-dependent peak velocity dispersion bias (red dashed) in Eq.~\eqref{eq:sig12}.}
\label{fig:evolvedpeaksv}
\end{figure}

%-------------------------------------------------------------------------
The mean motion of peaks can be understood based on their initial velocity statistics. The displacement in the Zel'dovich approximation is just the time integral of the velocity. In the previous section we assumed that the peaks move according to their initial velocities, where $\boldsymbol\Psi=\boldsymbol v/\mathcal{H}f$. Using this relation between displacements and velocities, we will continue working with displacements but note that our results are related to velocity statistics by a simple time dependent but scale-independent prefactor. Let us for this purpose consider the mean infall $v_{12}/\mathcal{H}f=\left \langle (\Psi_2-\Psi_1)(1+ \delta_\text{pk,1})(1+ \delta_\text{pk,2}) \right \rangle/(1+\xi_\text{pk})$, where $\Psi_i=v_i/\mathcal{H}f$ and $\delta_{\text{pk},i}$ are the displacements and peak densities at positions 1 and 2 respectively. This mean streaming is an important ingredient for redshift space distortion models (see e.g. \cite{Reid:2011to}). In a fashion similar to what was presented above for displaced peaks but taking $D_{+}=0$, we have
\be
(1+\xi_\text{pk})\frac{v_{12}}{\mathcal{H}f}=\frac{\left\langle (v_2-v_1)(1+\delta_\text{pk,1})(1+\delta_\text{pk,2})\right \rangle}{\mathcal H f} = -\frac{1}{\bar n_\text{pk}^2}\!\int\! \derd \boldsymbol X\; \boldsymbol X^{\rm T} \!\!\cdot\!  \mathbf{C}_{ {X}} ^{-1}  \!\!\cdot\! \mathbf{C}_{{\!X\!\Psi}} \!\!\cdot\! \,\boldsymbol u\; \frac{w(\boldsymbol X)}{\sqrt{(2\pi)^6 \det \mathbf{C}_{ {X}} }} \eh{-\frac12 \boldsymbol X^{\rm T}\!\!\cdot\! \mathbf{C}_{ {X}} ^{-1} \!\!\cdot\!\boldsymbol X}\; .
\ee
On large scales this quantity can be approximated by its linear (scale-dependent) bias expansion
\begin{equation}
\frac{v_{12}}{\mathcal{H}f}\approx -\left[2 b_{10} \left(\xi_{-1/2}-\frac{\sigma_{0}^2}{\sigma_1^2} \xi_{1/2}\right) +
 2b_{01} \left(\xi_{1/2} - \frac{\sigma_{0}^2}{\sigma_{1}^2} \xi_{3/2}\right)\right]\frac Q {|Q|}\; ,
 \label{eq:linvelbias}
\end{equation}
in agreement with 3-dimensional results \citep{desjacques/sheth:2010}.
By contrast, the mean infall in the local bias model reads $v_{12}\approx -2 b \xi_{-1/2} Q/|Q|$, where $\xi_{-1/2}$ is usually evaluated without an explicit smoothing scale.
In Fig.~\ref{fig:evolvedpeaksv} we show the mean relative velocity of the same sample of peaks considered for the evolution above, i.e., peaks of height $\bar \nu =3/2$ and $\Delta \nu=3/5$ in a $n=3$ power law density field. For the underlying matter distribution we have $v_{12}/(\mathcal{H}f)\approx -2\xi_{-1/2}Q /{|Q|}$. We clearly see that matter and peak mean streaming differ for scales $r<10 R_\star$. This deviation is captured by the scale-dependent peak velocity bias equation~\eqref{eq:linvelbias} down to $r\approx 7 R_\star$. Thus, the linear peak velocity bias starts to deviate at larger separations than the linear peak density bias. Note that the distance shown in this figure corresponds to the mean offset between the Eulerian separation and the Lagrangian separation in Fig.~\ref{fig:displcontrib}.
Let us finish our considerations about the peak dynamics by considering the velocity dispersion, $(\sigma_{12}/\mathcal{H}f)^2={\left\langle (\Psi_2-\Psi_1)^2(1+\delta_\text{pk,1})(1+\delta_\text{pk,2})\right \rangle}/(1+\xi_\text{pk})$ , which can be exactly calculated as
\be
(1+\xi_\text{pk})\frac{\sigma_{12}^{2}}{(\mathcal{H}f)^2}=\frac{1}{\bar n_\text{pk}^2}\!\int \!\derd \boldsymbol X\; \Bigl[\left(\boldsymbol X^{\rm T} \!\!\cdot\!  \mathbf{C}_{ {X}} ^{-1}  \!\!\cdot\! \mathbf{C}_{{\!X\!\Psi}} \!\!\cdot\! \,\boldsymbol u\right)^2+\boldsymbol u^{\rm T} \!\!\cdot\!  \mathbf{\Omega}_{\boldsymbol {\Psi}}^{-1} \!\!\cdot\! \, \boldsymbol u\Bigr]\; \frac{w(\boldsymbol X)}{\sqrt{(2\pi)^6 \det \mathbf{C}_{\boldsymbol {X}} }} \eh{-\frac12 \boldsymbol X^{\rm T}  \!\!\cdot\!  \mathbf{C}_{\boldsymbol {X}} ^{-1}  \!\!\cdot\! \boldsymbol X} \!.
\ee
For large separations -- or equivalently small correlation functions --, we get the following approximate expression
\be
\label{eq:sig12}
\frac{\sigma_{12}^{2}}{(\mathcal{H}f)^2}\approx 2 \sigma_{v,\text{pk}}^2 - 
 2 \left(\xi_{-1} - 2\frac{\sigma_{0}^2}{\sigma_1^2} \xi_{0} + \frac{\sigma_{0}^4}{\sigma_1^4} \xi_1 \right)\,.
\ee
This result could have been obtained using the peak density and velocity bias in
\be
\sigma_{12}^2\approx\left(\left\langle v_1^2\right \rangle+\left\langle v_2^2\right \rangle\right)-2\left \langle v_1 v_2\right \rangle\; .
\ee
Equation~(\ref{eq:sig12}) is to be compared with the expectation of the velocity dispersion in the local bias model which reads at leading order $2 \sigma_{-1}^2-2 \xi_{-1}$ evaluated without an explicit smoothing scale.

In this section we have derived an expression for calculating the evolved peak correlation function on all scales, assuming that the peaks evolve according to the Zel'dovich approximation. We have seen that non-perturbative effects persist even for the evolved field, which is partially due to the peak effects in the mean relative displacement and displacement dispersion.
%

%>>>>>>>>>>>>>>>>>>>>>>>>>>>>>>>>>>>>>>>>>>>>>>>>>>>>>>>>>>>>>>>>>>>>>>>>>>>>>>>>>>>>
%>>>>>>>>>>>>>>>>>>>>>>>>>>>>>>>>>>>>>>>>>>>>>>>>>>>>>>>>>>>>>>>>>>>>>>>>>>>>>>>>>>>>

%%%%%%%%%%%%%%%%%%%%%%%%%%%%%%%%%%%%%%%%%%%%%%%%%%%%%%%%%%%%%%%%%%%%%%%%%%
\section{Conclusions}
%%%%%%%%%%%%%%%%%%%%%%%%%%%%%%%%%%%%%%%%%%%%%%%%%%%%%%%%%%%%%%%%%%%%%%%%%%

\label{sec:conclusion}
Halo exclusion is an essential ingredient towards a realistic description of the two-halo term,
which encodes most of the cosmological information that can be extracted from the two-point 
correlation of biased tracers. Halo exclusion stems from the fact that one cannot find two
peaks of the mass density field (haloes) arbitrary close to each other or even overlap.
Until very recently however, this effect had been either completely ignored or crudely modelled by setting $\xi_\text{h}(r)=-1$ 
at separations $r$ less than the sum of the halo virial radii, while sticking to the linear bias 
approximation at larger distances.

In this paper, we have investigated this small-scale exclusion using a simple, well-motivated 
approximation: the clustering of Gaussian density peaks and, more general, critical points in 
one dimension. After studying the small-$r$ behaviour for various power-law spectra and its 
sensitivity to the peak height, we have shown that the two-point correlation function of 1D density 
peaks differs significantly from the crude, aforementioned prescription. We have also explored 
how peak exclusion affects the convergence of the perturbative bias expansion in real and Fourier 
space. Finally, we have included the displacement from the initial to final peak position according
to the Zel'dovich approximation to clarify how exclusion effects mix up with scale-dependencies
induced by the nonlinear gravitational evolution.

Our key findings can be summarized in the following points:
\begin{enumerate}
\item the correlation function of equal height peaks or critical points ($\Delta \nu=0$) asymptotes to a finite non-zero number at small scales
\item the correlation function of unequal height peaks or critical points ($\Delta \nu \neq 0$) deviates from the aforementioned case on small scales
and asymptotes to exactly $-1$, where a percent deviation from $\Delta \nu=0$ is typically reached for $r_{1\%}\approx \left(180 \Delta \nu\right)^{1/3}R_\star$; this scale is valid for a broad range of power spectra including projected $\Lambda$CDM
\item the local bias expansion fails to describe the scale-dependence starting at very large scales, the peak bias series including derivative operators fares better but its convergence in the $r\approx R_\star$-regime is very slow and it completely fails in the exclusion regime
\item time evolution enhances the large-scale clustering according to the well-known $b+D_{\!+}$ behaviour and leads to significant modifications in the exclusion regime where non-linearities reduce the small-scale clustering of peaks
\item the statistics of the relative displacement of peaks is within reach of the formalism and is shown to depart significantly from the mean streaming of the dark matter field, the validity of the peak bias expansion in this regime is reduced compared to the case of the density field.
\end{enumerate}

Even though our findings apply, strictly speaking, to the clustering of tracers in one-dimensional density fields, 
they provide useful insights into halo exclusion and its impact on the two-halo term in a realistic setting in three dimensions. It would certainly be interesting but computationally more challenging to study these effects for peaks in three dimensional density fields.
This  will be addressed in an upcoming paper.

%%%%%%%%%%%%%%%%%%%%%%%%%%%%%%%%%%%%%%%%%%%%%%%%%%%%%%%%%%%%%%%%%%%%%%%%%%
\section*{Acknowledgments}
%%%%%%%%%%%%%%%%%%%%%%%%%%%%%%%%%%%%%%%%%%%%%%%%%%%%%%%%%%%%%%%%%%%%%%%%%%
We thank D. Pogosyan for useful comments during the course of this work.
VD thanks the LABEX ``Institut de Lagrange de Paris''  for funding and acknowledges support
from the Swiss National Science Foundation. 
TB gratefully acknowledges support by the Institute for Advanced Study through the Corning 
Glass works foundation fund.
CP and SC thank the University of Geneva for funding and the community of 
 {\tt http://mathematica.stackexchange.com} for technical advice.
 This research is part of the Spin(e) (ANR-13-BS05-0005, \url{http://cosmicorigin.org}) 
 and the  Cosmo@NLO grants of the French Agence Nationale de la Recherche.

\bibliographystyle{mn2e}
\bibliography{references}

\appendix
%%%%%%%%%%%%%%%%%%%%%%%%%%%%%%%%%%%%%%%%%%%%%%%%%%%%%%%%%%%%%%%%%%%%%%%%%%
\section{Explicit expression for small radii}
%%%%%%%%%%%%%%%%%%%%%%%%%%%%%%%%%%%%%%%%%%%%%%%%%%%%%%%%%%%%%%%%%%%%%%%%%%
\label{sec:app}
At small radii, it has been shown in Section~\ref{sec:smallR} that the signed critical point correlation function can be expanded at small separations
\begin{equation}
\begin{split}
1+\xi_\text{crit}\left(\tilde r,\nu-\frac{\Delta \nu}{2},\nu+\frac{\Delta \nu}{2},n=0\right)=&\frac{\sqrt 3}{\left(\Delta \nu ^2-4 \nu^2\right)}
\biggl[
\alpha_{8}(\nu,\Delta\nu)\frac{ \Delta \nu ^2}{\tilde r^8 }
+
  \alpha_{6}(\nu,\Delta\nu) \frac{ \Delta \nu ^2}{\tilde r^6 }
   +
   \alpha_{4}(\nu,\Delta\nu)\frac{\Delta \nu ^2}{\tilde r^4 }
   +
 \alpha_{2}(\nu,\Delta\nu)  \frac{ \Delta \nu ^2}{5 \tilde r^2 } \\
   &+{\alpha_{0}(\nu,\Delta\nu)}{ }
   \biggr]
   \exp\left[\frac{\nu ^2}{4}+\frac{7 \Delta \nu ^2}{80}-\frac{9 \Delta \nu ^2}{5 \tilde r^2}+\frac{27
   \Delta \nu ^2}{\tilde r^4}-\frac{324 \Delta \nu ^2}{\tilde r^6}\right]+{\cal O}(\tilde r)\,,
\end{split}
\end{equation}
where the polynomials $\alpha_{2n}(\nu,\Delta\nu)$ for $n$ between 0 and 4 are given by
\begin{eqnarray}
\alpha_{0}(\nu,\Delta\nu)&=&\!-\!\frac{3 \nu ^2+8}{6}
\!+\!\frac{\Delta \nu ^2 \!\!\left(315 \nu ^8\!+1680 \nu ^6\!+6720
   \nu ^4\!-121856 \nu ^2\!+280576\right)}{1720320}
 \!+\!\frac{\Delta \nu ^4\!\! \left(3645 \nu ^6\!+5940 \nu ^4\!+55776 \nu ^2\!-409856\right)}{43008000}\nonumber\\
   &&+
   \frac{9 \Delta \nu ^6 \left(32805 \nu ^4-16200 \nu
   ^2+258976\right)}{20070400000}  
   +\frac{59049 \Delta \nu ^8 \left(27 \nu ^2-28\right)}{1404928000000}+
   \frac{129140163 \Delta \nu ^{10}}{3933798400000000},\nonumber
\\
\alpha_{2}(\nu,\Delta\nu)&=&\frac{3 \left(2048-320 \nu ^2-15 \nu ^6\right)}{1280}
-\frac{3 \Delta \nu ^2 \left(3645 \nu ^4-5760 \nu
   ^2+39232\right)}{896000}
-\frac{2187 \Delta \nu ^4 \left(81 \nu ^2-128\right)}{125440000}
-\frac{4782969 \Delta \nu ^6}{87808000000}\nonumber,
   \\
 \alpha_{4}(\nu,\Delta\nu)&=& \frac{9}{16} \left(9 \nu ^4-24 \nu ^2+128\right) 
+\frac{81 \Delta \nu ^2 \left(81 \nu ^2-172\right)}{5600}+ \nonumber
\frac{531441 \Delta \nu ^4}{7840000} ,  \\
 \alpha_{6}(\nu,\Delta\nu)&=&162 \left(8-3 \nu ^2\right)\nonumber
-\frac{19683 \Delta \nu ^2}{350} ,  \\\nonumber
 \alpha_{8}(\nu,\Delta\nu)&=&23328\,.
\end{eqnarray}
When $\Delta \nu>0$, the signed critical point correlation function tends to $-1$ at small separations and therefore shows a clear exclusion zone while for $\Delta\nu=0$, it tends to a constant different from $-1$ and given by 
\begin{equation}
1+\xi_\text{crit}\left(\tilde r,\nu,\nu,n=0\right)=\sqrt 3\left(\frac{1}{8}+\frac {1}{3\nu^{2}}\right)   \exp\left[\frac{\nu ^2}{4}\right]+{\cal O}(\tilde r)\,.
\end{equation}
It has to be noted that this constant is negative only for $\nu$ roughly between 1 and 2 for $n=0$ power-law power spectra.

%%%%%%%%%%%%%%%%%%%%%%%%%%%%%%%%%%%%%%%%%%%%%%%%%%%%%%%%%%%%%%%%%%%%%%%%%%
\section{Covariance Matrix}
%%%%%%%%%%%%%%%%%%%%%%%%%%%%%%%%%%%%%%%%%%%%%%%%%%%%%%%%%%%%%%%%%%%%%%%%%%
\label{app:cov}
The covariance matrice of $\boldsymbol X$ and $\boldsymbol \Psi$ defined in Section~\ref{sec:zeldo} are given by
\be
\mathbf{C}_{ {X}}=\left(
\begin{array}{cccccc}
 \sigma _0^2 & 0 & -\sigma _1^2 & \xi _0 & -\xi _{\frac{1}{2}} & -\xi _1 \\
 0 & \sigma _1^2 & 0 & \xi _{\frac{1}{2}} & \xi _1 & -\xi _{\frac{3}{2}} \\
 -\sigma _1^2 & 0 & \sigma _2^2 & -\xi _1 & \xi _{\frac{3}{2}} & \xi _2 \\
 \xi _0 & \xi _{\frac{1}{2}} & -\xi _1 & \sigma _0^2 & 0 & -\sigma _1^2 \\
 -\xi _{\frac{1}{2}} & \xi _1 & \xi _{\frac{3}{2}} & 0 & \sigma _1^2 & 0 \\
 -\xi _1 & -\xi _{\frac{3}{2}} & \xi _2 & -\sigma _1^2 & 0 & \sigma _2^2 \\
\end{array}
\right)\,,
\ee
\be
\mathbf{C}_{{X\Psi}}^{\rm T}=\left(
\begin{array}{cccccc}
 0 & \sigma _0^2 & 0 & \xi _{-\frac{1}{2}} & \xi _0 & -\xi _{\frac{1}{2}} \\
 -\xi _{-\frac{1}{2}} & \xi _0 & \xi _{\frac{1}{2}} & 0 & \sigma _0^2 & 0 \\
\end{array}
\right)\,,
\quad
{\rm and}
\quad
\mathbf{C}_{ {\Psi}}=\left(
\begin{array}{cc}
 \sigma _{-1}^2 & \xi _{-1} \\
 \xi _{-1} & \sigma _{-1}^2 \\
\end{array}
\right)\,.
\ee

\end{document}